\documentclass[a4paper,fleqn]{cas-sc}

\usepackage[numbers]{natbib}
\usepackage{graphicx} % Required for inserting images
\usepackage{float}
\usepackage{url, hyperref}
\usepackage{rotating}

\usepackage{lineno}
\usepackage{subfig,subfloat} % packages for subfigure labeling '(a)...(b)'. See Fig. 17 for example. (Timo)
\usepackage{upgreek} % package for upright micron symbol '\upmu' (Timo)
\usepackage{xcolor}  % package for colored text
\usepackage{siunitx} % SI unit package
\usepackage{svrsymbols} %Table 336: svrsymbols Physics Ideograms
\usepackage[utf8]{inputenc} % Required for inputting international characters
\usepackage[T1]{fontenc} % Output font encoding for international characters

\def\tsc#1{\csdef{#1}{\textsc{\lowercase{#1}}\xspace}}
\tsc{WGM}
\tsc{QE}
\begin{document}
\let\WriteBookmarks\relax
% \def\floatpagepagefraction{1}
% \def\textpagefraction{.001}

% Short title
\shorttitle{Impact of Environmental Stress on LGAD Sensors' Response}    

% Short author
\shortauthors{M. H. {Mohamed Farook} et al.}  

% Main title of the paper
\title [mode = title]{Impact of Environmental Stress on Low Gain Avalanche Diode Sensors' Response}  

% Title footnote mark
% eg: \tnotemark[1]
%AT \tnotemark[1] 

% Title footnote 1.
% eg: \tnotetext[1]{Title footnote text}
%AT \tnotetext[1]{} 

% First author
%
% Options: Use if required
% eg: \author[1,3]{Author Name}[type=editor,
%       style=chinese,
%       auid=000,
%       bioid=1,
%       prefix=Sir,
%       orcid=0000-0000-0000-0000,
%       facebook=<facebook id>,
%       twitter=<twitter id>,
%       linkedin=<linkedin id>,
%       gplus=<gplus id>]

%AT: \author[1]{}%[<options>]
\author[newmexico]{Mohamed Hijas {Mohamed Farook}}
\author[brown]{Gaetano Barone}
\author[physics]{Diallo Boye}
\author[physics]{Gabriele D'Amen}
\author[instrumentation]{Gabriele Giacomini}
\author[TTU]{Timo Peltola}
\author[brown]{Jennifer Roloff}
\author[physics]{Enrico Rossi}
\author[brown]{Trevor Russell}
\author[newmexico]{Sally Seidel}
\author[physics]{Alessandro Tricoli}\cormark[1]
\cortext[1]{Corresponding author}% Email id of the first author
\ead{atricoli@bnl.gov}
% Credit authorship
%\credit{0000-0002-8224-6105}
\author[browncs]{Lixing Wang}
\address[newmexico]{Department of Physics and Astronomy, University of New Mexico, Albuquerque, 87106 New Mexico, United States}
\address[brown]{Department of Physics, Brown University, Providence, 02912 Rhode Island, United States}
\address[physics]{Physics Department, Brookhaven National Laboratory, Upton, 11973 New York, United States}
\address[instrumentation]{Instrumentation Department, Brookhaven National Laboratory, Upton, 11973 New York, United States}
\address[TTU]{Department of Physics and Astronomy, Texas Tech University, Lubbock, 79409 Texas, United States}
\address[browncs]{Department of Computer Science, Brown University, Providence, 02912 Rhode Island, United States}

% Corresponding author indication
% Footnote of the first author
%AT \fnmark[1]
% URL of the first author
%AT \ead[url]{}

% Credit authorship
% eg: \credit{Conceptualization of this study, Methodology, Software}
%AT \credit{}

% Address/affiliation
%AT \affiliation[newmexico]{organization={Department of Physics and Astronomy, University of New Mexico},
%AT             addressline={}, 
%AT             city={Albuquerque},
%          citysep={}, % Uncomment if no comma needed between city and postcode
%AT             postcode={87106}, 
%AT             state={New Mexico},
%AT             country={United States}}

%AT \author[2]{}%[]
% Footnote of the second author
%AT \fnmark[2]
% Email id of the second author
%AT \ead{}
% URL of the second author
%AT \ead[url]{}

% Address/affiliation
%AT \affiliation[2]{organization={},
%AT             addressline={}, 
%AT             city={},
%          citysep={}, % Uncomment if no comma needed between city and postcode
%AT             postcode={}, 
%AT             state={},
%AT             country={}}

% Corresponding author text
%AT \cortext[1]{Corresponding author}

% Footnote text
%AT \fntext[1]{}

% For a title note without a number/mark
%\nonumnote{}

% Here goes the abstract
%Here goes the abstract \nocite{*}%% Remove this line from your manuscript.
% \include{sections/abstract}
%AT \linenumbers
\begin{abstract}
Low Gain Avalanche Diodes or LGADs are silicon sensors capable of achieving excellent timing resolution due to their characteristic internal gain. Detectors based on LGAD technology play a crucial role in High Energy and Nuclear Physics experiments, among other applications. However, their performance is affected by environmental factors such as temperature, humidity, and storage conditions. A systematic evaluation of the response of LGAD sensors as a function of these environmental parameters is therefore of essential importance for any application. LGAD sensors fabricated at the Brookhaven National Laboratory are characterized and stress-tested against various operating conditions, such as rapid temperature and humidity changes. Dedicated and detailed simulations are used to interpret the experimental results. 
\end{abstract}

% Use if graphical abstract is present
%\begin{graphicalabstract}
%\includegraphics{}
%\end{graphicalabstract}

% Research highlights
%AT \begin{highlights}
%AT \item 
%AT \item 
%AT \item 
%AT \end{highlights}

% Keywords

\maketitle

% Main text
%AT: \section{}\label{}
\section{Introduction}

Silicon devices with an internal gain layer, such as Low Gain Avalanche Diodes (LGADs)~\cite{hartmut,Giacomini_2019_LGAD}, are quickly becoming the technology of choice for multiple applications in High Energy Physics~\cite{CMS:2667167,Collaboration:2623663}, Nuclear Physics~\cite{AbdulKhalek:2021gbh}, and other fields~\cite{micronlgad,pioneercollaboration2022testingleptonflavoruniversality,instruments5040040}. These sensors have shown the ability to achieve time resolution on the order of 30 ps or less~\cite{Heller_2022}.
A variant of the technology, the AC-coupled LGAD (AC-LGAD)~\cite{ACLGADprocess}, allows precise particle tracking information, at  the micron-level~\cite{Heller_2022}, in addition to precise timing. %in line or exceeding the requirements for multiple application in high-energy physics experiments and nuclear experiments, space applications, and others.
This performance by LGADs and AC-LGADs is achieved through the implementation of a  low to moderate gain (10-100) as a consequence of the high electric field present within the semiconductor material that causes impact ionization of the drifting electrons. The design of these structures that exploits this charge multiplication effect that causes high signal-to-noise ratio, together with a short rise time that depends on the active thickness of the device, leads to a silicon detector capable of time measurements with high resolution.  In addition, AC-LGAD structures allow for pixelation of the top metal layer, which enables precise resolution of particle hit position as in a pixel detector. 

These novel sensors are expected to be used in a variety of applications, for instance in space science, radiation therapy, fusion plasma diagnostics, light source experiments, ion spectroscopy among others. The sensors will be operated in different environmental conditions, according to the different experimental setups, and will be subject to different handling procedures during testing, assembly, transport and operations, therefore a thorough quality assurance of their long-term robustness under different, varying and  extreme conditions is necessary. Thus, LGADs and AC-LGADs require precise understanding and precise modeling of their response as a function of several environmental parameters, such as temperature, humidity, and storage conditions. 
%For example, the phenomenon of phonon scattering plays a central role in avalanche multiplication at higher temperatures, where phonon scattering becomes prominent. At high temperatures, charge carriers tend to lose their energy as they travel through the multiplication region and therefore require longer paths before they acquire sufficient energy from the field to effect impact ionization. Thus, at high temperatures, the avalanche multiplication process is expected to weaken. This has ramifications on the sensor noise, gain, and breakdown voltage. Other environmental parameters can negatively affect the sensor gain and operational conditions, such as humidity and build-up static charge. 

Both LGADs and AC-LGADs are routinely designed and fabricated at Brookhaven National Laboratory (BNL)~\cite{Giacomini_2019_LGAD,ACLGADprocess,DUTTA2025170224} where they can be stored, tested, and operated in strictly controlled environments, such as in controlled temperature and humidity conditions. This paper reports the impact on the electrical performance of LGADs and AC-LGADs produced at BNL under the sudden and repeated change of environmental parameters.

The paper is organized as follows: Section \ref{sec:theory} introduces the expected behavior of silicon devices under different environmental conditions; Section \ref{sec:setup} details the devices under study and the instruments used for their characterization; Section \ref{sec:characterization} presents the data obtained during the environmental characterization of the devices, while Section \ref{sec:analysis} discusses the experimental results; finally, Section \ref{sec:simulations} presents the simulation and its results compared to the experimental results, while Section \ref{sec:conclusions} draws conclusions.
\section{Silicon sensor response to temperature and humidity}\label{sec:theory}
The ability of silicon detectors, as semiconductor crystals, to form an electric signal in response to the passage of a particle depends on the mechanisms that regulate charge formation and transport within semiconductor materials. While charge carriers within silicon crystals obey well understood rules, small changes in their formation and transport properties can lead to significant differences in the final performance of a detector. 
Temperature and humidity conditions might impact the electrical performance of silicon sensors by affecting %the mobility of the charge carriers, 
the leakage current. %and the carrier charge multiplication and collection. 

\subsection{Impact of temperature on leakage current}
\label{sec:theory_leakage}

%from: https://pubs.aip.org/aip/adv/article/10/12/125301/994058/Leakage-current-analysis-of-silicon-diode-with

The leakage current of a reverse-biased silicon sensor is the sum of bulk and surface contributions. Within the $p$–$n$ junction model, the bulk contribution in a diode under a reverse bias condition consists of a generation current, due to thermal generation of carriers via deep energy levels in the depleted volume, and a diffusion current, due to minority carriers generated in the undepleted regions within a diffusion length of the depletion edge.
The two components have different temperature dependencies, as they are proportional to the intrinsic carrier concentration, $n_i$, and its square, $n_i^2$, respectively, where the temperature dependence of $n_i$ is as follows~\cite{Sze2007}:
%The generation and diffusion leakage currents have slightly different temperature dependencies, as they are proportional to the intrinsic carrier concentration, $n_i$, and its square, $n_i^2$, respectively, where the temperature dependence of $n_i$ is as follows~\cite{Sze2007}:
%
\begin{equation}
    n_i \propto T^{3/2} \, e^{-E_g / 2kT},
\end{equation}
with $E_g$ being the energy bandgap (1.12 eV at room temperature), $T$~the temperature in Kelvin unless differently specified, and $k$~the Boltzmann constant.
Including the temperature dependence of the generation lifetime (through the carrier thermal velocity) and of the bandgap, and neglecting the weak temperature dependence of the diffusion parameters, the two components scale as~\cite{Sze2007,Chilingarov_2013}
\begin{equation}
   I_\text{gen} \propto T^2 \exp\!\left(-\frac{E_\text{eff}}{2kT}\right), \qquad I_\text{diff} \propto T^{3} \exp\!\left(-\frac{E_\text{eff}}{kT}\right), \tag{2}
\end{equation}
where $E_\text{eff} \approx 1.21$ eV is the effective bandgap, which approximately corresponds to the linear extrapolation of the temperature-dependent bandgap to $T = 0$ K~\cite{Chilingarov_2013}; thus  the leakage current rises steeply with increasing temperature.
%AT: added
 At temperatures below about 100 $^{\circ}$C, the leakage current is expected to be dominated by the generation current component, with an activation energy $E_a \approx E_\text{eff}/2 \approx 0.6$ eV,
 %$E_a \approx E_g/2 \approx 0.5 - 0.6$ eV, 
 whereas at higher temperatures the diffusion component is expected to dominate, with an activation energy $E_a \approx E_\text{eff} \approx 1.2$ eV~\cite{ActivationEnergy}. Surface contributions are discussed in Sec.~\ref{sec:theory_humidity}.

In LGADs, electrons generated in the depleted bulk drift through the gain layer and are multiplied by impact ionization, so the bulk generation current measured at the pad is enhanced by a gain factor. Since the impact ionization coefficients increase with decreasing temperature~\cite{UniBo2005,Massey2006}, the gain increases and the breakdown voltage decreases as the sensor is cooled. The temperature dependence of the LGAD leakage current therefore results from the competition between the decrease of generation current and the increase of gain, and deviates from the pure $n_i(T)$ scaling of a no-gain diode, particularly near breakdown.
%AT

As the leakage current contributes directly to the noise of the detector, maintaining a stable, well-controlled operating temperature is critical for achieving optimal signal-to-noise performance.

%%%%%%%%%%%%%%%%%%%%%%%%%%%%%%%%%%%%%%%%%%%%%%%%%%%%%
\subsection{Impact of humidity on leakage current}
\label{sec:theory_humidity}

The surface of a silicon detector is usually covered by a protective thin layer of silicon dioxide (SiO$_2$), which acts as a passivation layer. The oxide contains fixed positive charges, of density of order $\sim 10^{11}$~cm$^{-2}$~\cite{Poehlsen2013time}, located close to the Si--SiO$_2$ interface, while the interface itself hosts electrically active traps~\cite{Sze2007}. Besides the bulk components discussed in Sec.~\ref{sec:theory_leakage}, the leakage current includes surface contributions originating at the Si--SiO$_2$ interface.
The dominant one is the surface generation current: thermal generation via interface traps in regions where the interface is depleted~\cite{Sze2007}. Like the bulk generation current, it scales with $n_i$. Unlike it, carriers generated at the interface outside the gain-layer area are collected by the nearest electrode without traversing the gain layer, and are therefore not multiplied.

Humidity affects the surface current through two distinct mechanisms: modification of the surface charge and ohmic conduction. 

A) {\it Modification of the surface charge}: in the $p$-type epitaxial layer of the sensors studied here, under dry or vacuum conditions the positive oxide charge inverts the surface, producing an electron inversion layer beneath the oxide. %the effective charge collection can be altered by the surface conditions at the Si--SiO$_2$ interface, due to humidity. 
Exposure to moisture deposits negative charge on the outer oxide surface, 
%which can overcompensate the intrinsic positive charges. 
modifying the electrostatic boundary conditions at the interface~\cite{Poehlsen2013time,Poehlsen2013charges}\footnote{The studies in Refs.~\cite{Poehlsen2013time,Poehlsen2013charges} were performed on sensors with n-type bulk; the same mechanism applies to p-type bulk with the roles of electrons and holes exchanged.}. This charge compensates the positive oxide charge, depleting the surface, or, if it overcompensates it, induces a hole accumulation layer.
Since interface traps generate carriers most efficiently when the interface is depleted, whereas generation is largely suppressed under inversion or accumulation, the humidity-induced depletion of the interface increases the surface generation current~\cite{Sze2007,Ranjan2002}. The change of the surface layers also affects the local electric field, in particular at the edge of the $n^{+}$ electrode, the charge collection efficiency of carriers generated near the surface, and the inter-electrode isolation of segmented structures. The resulting surface leakage current can increase by an order of magnitude under high humidity conditions~\cite{Ranjan2002}.

B) {\it Ohmic conduction}: adsorbed moisture forms a thin conducting water film whose sheet resistance decreases with increasing relative humidity~\cite{Poehlsen2013time}, providing an additional, approximately ohmic, surface leakage path,
%Furthermore, the redistribution of surface ions is governed by the sheet resistance of the surface, which itself depends strongly on relative humidity~\cite{Poehlsen2013time}.
%This conductive layer provides a path for surface leakage currents
which can increase the overall leakage current of the device and degrade the signal-to-noise ratio.
At high relative humidity, the dielectric strength of weak spots in the passivation layer may be reduced\footnote{Water molecules can accumulate in microscopic fabrication defects of the passivation layer, such as surface pits, pores, crevices, and residual chemical impurities, and form a conductive path under the applied electric field. Dielectric strength refers to the maximum electric field the insulating layer can reach before breakdown.}, creating localized conducting channels and causing premature breakdown~\cite{Karrevula2023}.

One of the objectives of the paper is to study these effects in the LGADs fabricated at BNL, as the severity of these effects depends on both the relative humidity and the quality and thickness of the surface passivation, with well-passivated detectors showing significantly more stable performance under varying environmental conditions~\cite{Ranjan2002}.

%\subsection{Impact of humidity on leakage current and charge collection}
%%%%%%%%%%%%%%%%%%%%%%%%%%%%%%%%%%%%%%%%%%%%%%%%%%%
\section{Experimental setup}\label{sec:setup}
%%%%%%%%%%%%%%%%%%%%%%%%%%%%%%%%%%%%%%%%%%%%%%%

Several LGADs and AC-LGADs fabricated by BNL were tested under different experimental conditions. The tests comprised measurements of current as a function of bias voltage (IV scans) and the environmental conditions were kept under control in a climate chamber, as detailed in the following.

%%%%%%%%%%%%%%%%%%%%%%
\subsection{Silicon sensors}
\begin{figure}[!htbp]
    \centering
   \includegraphics[width=.45\linewidth]{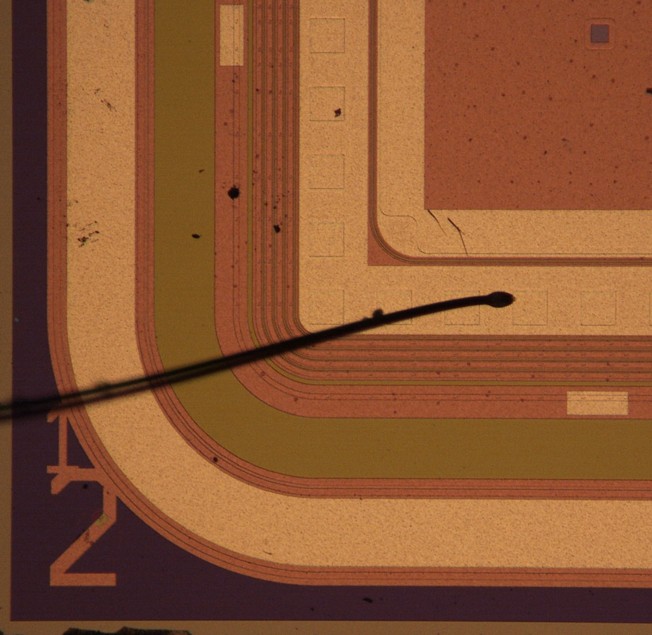}
   \includegraphics[width=.46\linewidth]{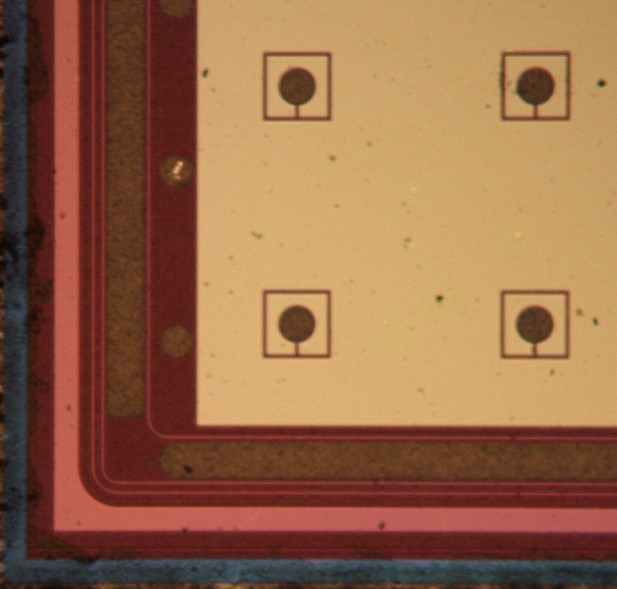}
    \caption{Photos of two sensors under test: the single-pad LGAD - same layout for both W3045 and W3058 wafers - (left), and the pixelated AC-LGAD from wafer W3081 (right).
    \label{fig:sensors}}
\end{figure}

Two LGAD and one AC-LGAD sensors were tested. In all sensors, the $p^+$ (also known as gain) layer was implanted with the same energy, while at slightly different doses to adjust the breakdown voltage. This results in a depletion voltage of the gain layer of about 20-25 V. The peak of the gain layer doping profile is at about 0.9 $\mu \text{m}$ from the silicon surface. The full depletion voltage (V$_\textrm{FD}$) of the bulk is in the order of 1 V above the gain layer depletion voltage, due to the high resistivity used in the epitaxial substrate. All sensors have an active area of 1.3 x 1.3 mm$^{2}$. The LGAD sensors feature a single pad whereas the AC-LGAD is pixelated. The two LGAD sensors have an active epitaxial thickness of 50 $\mu \text{m}$ and were produced in two different silicon wafers, named W3045 and W3058. The LGAD from wafer W3045 comes from an early production at BNL, which does not feature a $p^+$  termination at the scribeline region and does not have a polymide or low-temperature oxide passivation on top of the thermal oxide, while the LGAD from wafer W3058 is from a more recent production and features $\text{Si}_3\text{N}_4/\text{SiO}_2$ passivation and improved contacts to the metal electrodes.
%AT: the following LGADs were used at CERN and not at BNL:
%Three LGADs have an active thickness of 20 $\mu \text{m}$ and were produced in the same wafer, namely W3076.  \textcolor{red}{(passivation, termination etc.?)}\textcolor{purple}{(3076 is not analyzed in the same way so I am not sure why it is relevant? But one would need to ask Gabriele about the passivation/termination, GG: w3076 is a DC LGAD, no passivation and no p-type termination. do we need to mention this device?).}
The AC-LGAD has an active epitaxial  thickness of 20 $\mu \text{m}$ and was produced in wafer W3081. In the following, the sensors are identified on the basis of their wafer identification code. Figure~\ref{fig:sensors} shows the three sensors.

%\begin{figure}[!htbp]
%    \centering
%   \includegraphics[width=.3\linewidth]{figures/setup/LGAD_W3045.png}
%   \includegraphics[width=.3\linewidth]{figures/setup/LGAD_W3058.png}
%   \includegraphics[width=.3\linewidth]{figures/setup/ACLGAD_W3081.png}
%    \caption{Photos of the three sensors: the single-pad LGAD from wafer W3045 (left), the single-pad LGAD from wafer W3058 (center), and the pixelated AC-LGAD from wafer W3081 (right).
%    \label{fig:sensors}}
%\end{figure}
%%%%%%%%%%%%%%%%%%%%%%%%%%%%
\subsection{Climate chamber and readout system}
\begin{figure}[htbp]
    \centering
    \includegraphics[width=.8\linewidth]{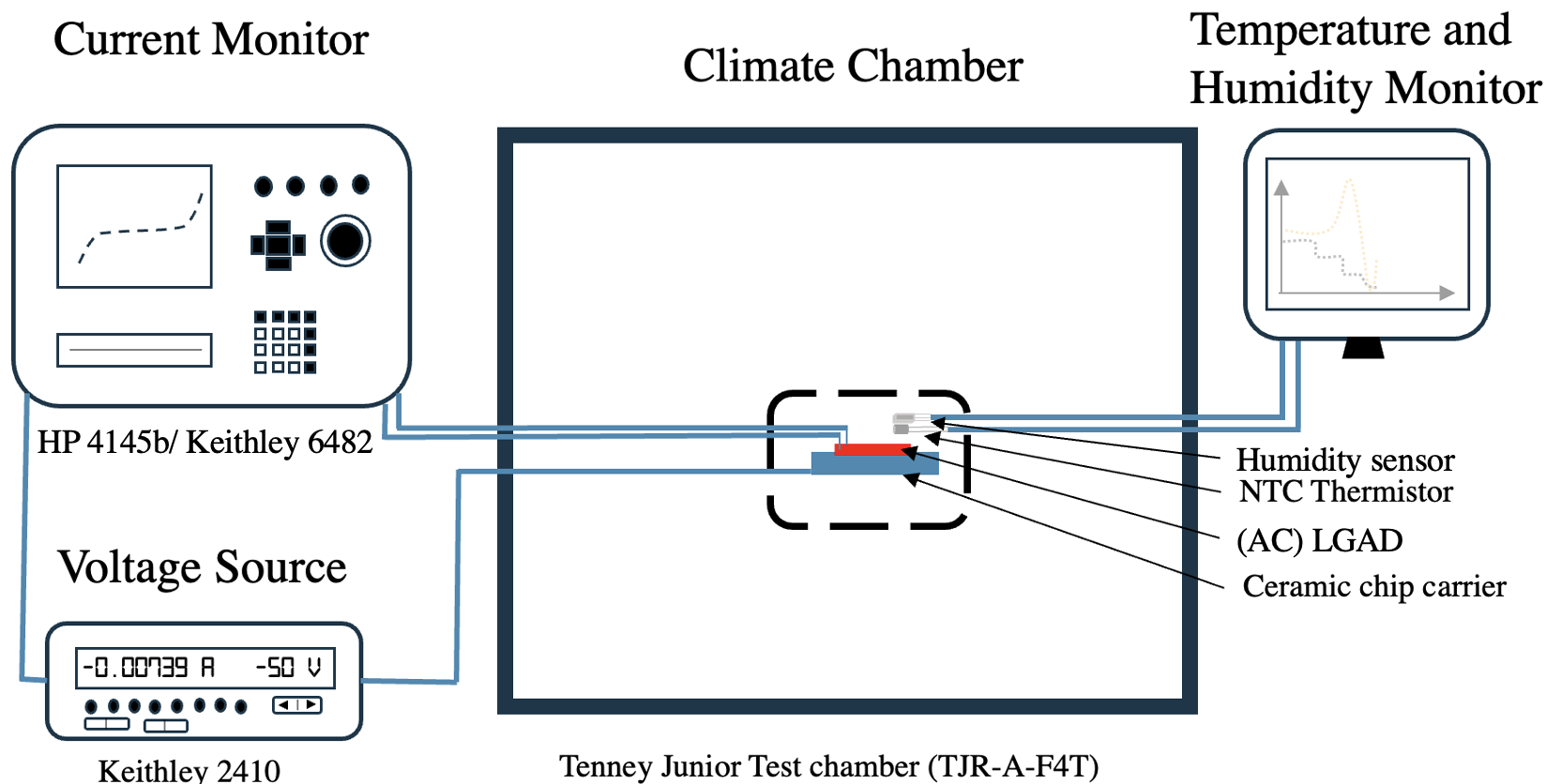}
    \caption{Schematic representation of the test setup. It shows the climate chamber that contains a sensor, mounted on a ceramic carrier board, as well as a thermistor and humidity sensor that are monitored by a temperature and humidity monitoring system. The design also shows the voltage source and the current measurement system.}
    \label{fig:testSetup}
\end{figure}

 A schematic drawing of the experimental setup used to vary the environmental conditions in which the sensors were tested is shown in Figure~\ref{fig:testSetup}. The LGAD and AC-LGAD sensors were mounted on individual  ceramic carrier boards, which allowed the sensors to be biased via a voltage source and their leakage currents read out using a current monitor. 
 The bias voltage is applied to a sensor by means of a Keithley 2410 SourceMeter~\cite{instr:keithley2410}, which also monitors its total leakage current. Leakage currents from the readout electrode and the guard ring of a sensor are also separately monitored using a high-precision HP 4145B semiconductor parameter analyzer~\cite{instr:HP4145b} (for LGADs) or the Keithley 6482 Dual-Channel Picoammeter (for the AC-LGAD)~\cite{Keithley6482}.
For the AC-LGADs the leakage current was measured at the DC-contact that is the electrode that surrounds the whole active area of the silicon sensors and collects the DC-current.
 The sensor pad and guard ring leakage currents of the AC-LGAD were measured using the Keithley 6482 Dual-Channel Picoammeter, which was sensitive in the 100 pA range, as compared to 1 pA for the HP 4145B Semiconductor Parameter Analyzer used for the LGADs.
 
\begin{figure}[htbp]
    \centering
    \includegraphics[width=.55\linewidth]{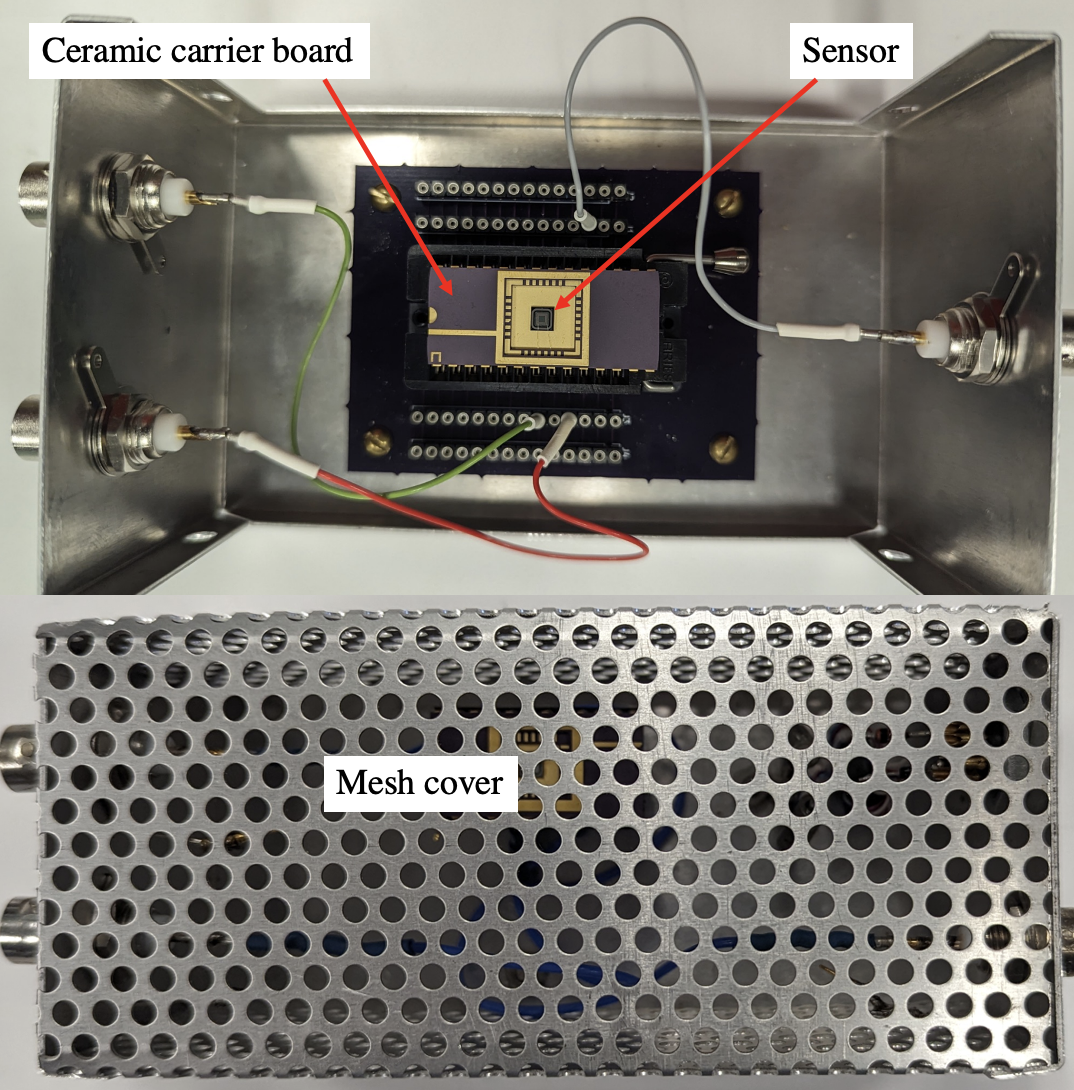}
    \caption{Photo of a sensor mounted on a ceramic test carrier.}
    \label{fig:metalbox}
\end{figure}

Temperature and humidity are monitored via a negative temperature coefficient (NTC) thermistor, mounted on the carrier board, and a humidity/temperature sensor of the Sensirion SHT7x series~\cite{sensirion2011sht7x}, which operates in the temperature range of -40 to +124 ${^\circ}$C. The NTC sensor, used to measure the temperature near the sensor, has a nominal uncertainty of $\pm$0.2~$^{\circ}$C, while the Sensirion SHT7x, used to report the relative humidity, has a nominal uncertainty of $1.8\%$. Those sensors were read out by a custom-made system based on an Arduino Uno~\cite{arduino_uno}, which was controlled by a Python program that was also used for real-time graphing. 
%pin-type relative humidity and temperature sensors. The device uses capacitive sensor element for humidity measurement and a band-gap sensor for temperature. .
The system of the carrier board, sensor, thermistor and humidity sensor is enclosed in a metal box with a mesh of fine holes drilled on three of its walls that serves as a Faraday cage to prevent electrical noise and allows air circulation, see Fig.~\ref{fig:metalbox}.
The metal box is placed inside a TJR-A-F4T climate chamber~\cite{climateChamber:TJR}, which is constantly supplied with $\text{N}_2$ to control the humidity levels inside the chamber and prevent condensation or moisture on the sensors. 
%An external gaseous $N_2$ supply is used to purge the chamber of moisture and to prevent condensation on the sensors.
%
An airflow system of elliptical/vertical-down type  is used in the climate chamber for air circulation. The chamber is heated by recirculating the air through open-air Nichrome wire heater elements, while cooling is achieved in the chamber by recirculating the air through a refrigerated finned cooling coil.
The ambient chamber can be operated within the temperature range from -68 to +180~$^{\circ}$C, with an uncertainty of $\pm$1~$^{\circ}$C. The temperature inside the chamber was monitored at the air inlet using a \SI{100}{\ohm} Platinum RTD sensor. 
 Temperature and humidity were monitored continuously to ensure that the sensor temperature was kept above the dew point, by controlling the delivered N$_2$ flux. 
%%%%%%%%%%%%%%%%%%%%%%%%%%%%%%%%%%%%%%%%%%%%
\section{Electrical characterization}\label{sec:characterization}
%%%%%%%%%%%%%%%%%%%%%%%%%%%%%%%%%%%%%%%%%%%%%%%%%%
Electrical parameters of the silicon sensors were characterized by performing IV scans under different temperature and humidity conditions inside the climate chamber.
The leakage currents were measured at the sensor pad for the LGADs; at the DC-contact for the AC-LGADs; and at the guard ring for all sensors, as a function of the sensor bias voltage (V$_{\mathrm{bias}}$).
During each IV scan, the bias voltage was ramped up in steps of 1 or 2~V starting from 0~V. The voltage ramp-up was stopped when the pad current reached a compliance value of 10~$\mu$A, to avoid damaging the silicon sensor. Once the sensor reached compliance (indicating an avalanche breakdown), the temperature and humidity were changed based on a predefined pattern, as described in the following; a new IV scan was performed after the temperature and the humidity of the system had reached a new equilibrium.

%%%%%%%%%%%%%%%%%%%%
\begin{table}[h!] \setlength{\tabcolsep}{2.5pt}
    \caption{Temperature ranges, steps, and V$_{\mathrm{bias}}$ for the thermal cycle programs performed on the silicon sensors, as well as the information about whether an IV scan was performed or not in a given program.}
    \label{tab:thermalCycles}
    \centering
    \begin{tabular}{c|c|c|c|c}
    \hline\hline
         Program    & Temp. Range [$^{\circ}$C] & Temp. Steps [$^{\circ}$C]  & V$_{\mathrm{bias}}$ at ramping [V]  & IV Scan \\\hline\hline
         Day        & -60 to +120     & +20               &               & \\
                    & +120 to -60     & -180              &  -50                  & At each Temp. Step\\
                    & -60 to +120     & +180              &                   & \\
                    & +120 to -60     & -20               &                   & \\ \hline
         Night      & +21             & None             & -150              & No scan \\ \hline
         Weekend    & +10 to +120     & $\pm20$               & OFF              & No scan\\\hline\hline
    \end{tabular}
    \label{tab:programs}
\end{table}
%%%%%%%%%%%%%%%%%%%%%%

\subsection{Thermal Cycles}
\label{sec:thremal_cycling}
The temperature inside the climate chamber was cycled to test the electrical performance and the long-time reliability of the sensors under different environmental conditions. Different thermal cycling programs were used during day, night, and weekends, as described in Table~\ref{tab:thermalCycles}. Each thermal cycling program was optimized for a different class of tests: the \textit{Day Program} was optimized for IV scans at varying temperatures, the \textit{Night Program} aimed to test the stability of the electrical performance under prolonged stable conditions, and the \textit{Weekend Program} was designed to test the robustness of the sensors under repeated thermal shocks when left unbiased.

\begin{figure}[t!]
    \centering
    \includegraphics[width=.55\linewidth]{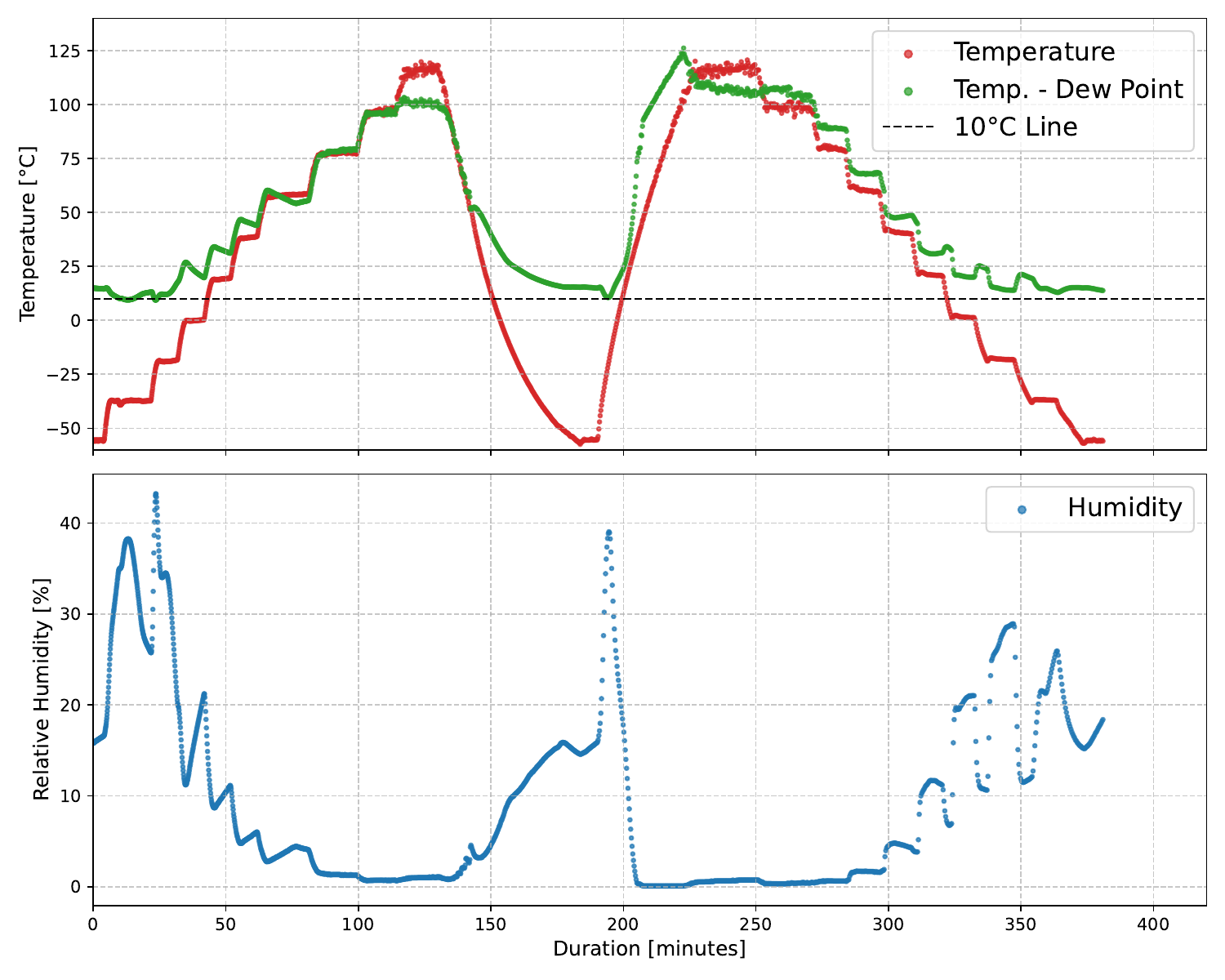}
    \caption{Temperature, dew point and humidity variations as a function of time in a typical Day Program. The variation of the dew point is graphed as the difference between the measured temperature and the dew point. The horizontal line marks the 10 ${^\circ}$C difference between temperature and dew point that is used as the safety margin of operations. The relative humidity and dew points were controlled by adjusting the N$_2$ flow.}
    \label{fig:day_program}
\end{figure}

The \textit{Day Program} allowed for electrical characterization to be performed during the thermal cycle. Temperature was ramped up from -60 to +120 $^{\circ}\text{C}$ in 20 $^{\circ}\text{C}$ steps, with an average rate of 1.81 $^{\circ}\text{C/minute}$.
After ramping to +120~$^{\circ}\text{C}$, the temperature was then rapidly cycled back to -60 $ ^{\circ}\text{C}$ and then to +120 $ ^{\circ}\text{C}$ again.
Next, the temperature was ramped down to -60 $^{\circ}\text{C}$ in 20 $^{\circ}\text{C}$ steps. The IV scans were taken at each temperature step. 
On average it took about 5.5 minutes to reach thermal equilibrium, so electrical tests were taken 10 minutes after the temperature had reached equilibrium. An IV scan took approximately 2 minutes, and for safety, during an IV scan, the thermal cycling program was paused and then resumed as soon as the scan was completed.
While the temperature was changing, the bias voltage of the silicon sensors was maintained at $-50$ V. Figure~\ref{fig:day_program} shows the temperature and relative humidity variations as a function of time in the Day Program.
\begin{figure}[h!]
    \centering
    \includegraphics[width=.55\linewidth]{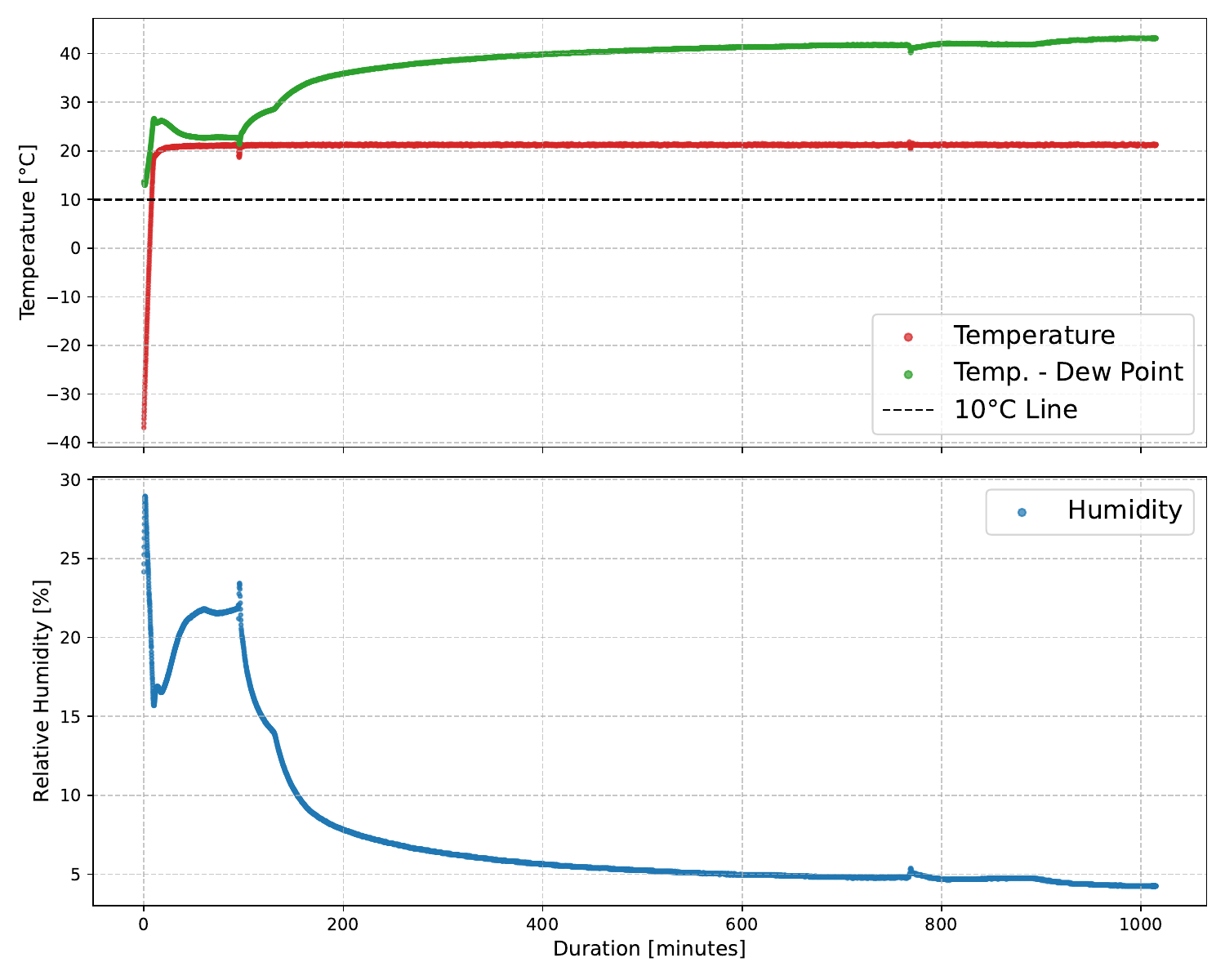}
    \caption{Temperature, dew point and humidity variations as a function of time in a typical Night Program. See Fig. \ref{fig:day_program} for the description.}
    \label{fig:overnight_program}
\end{figure}
\textbf{\begin{figure}[h!]
    \centering
    \includegraphics[width=.55\linewidth]{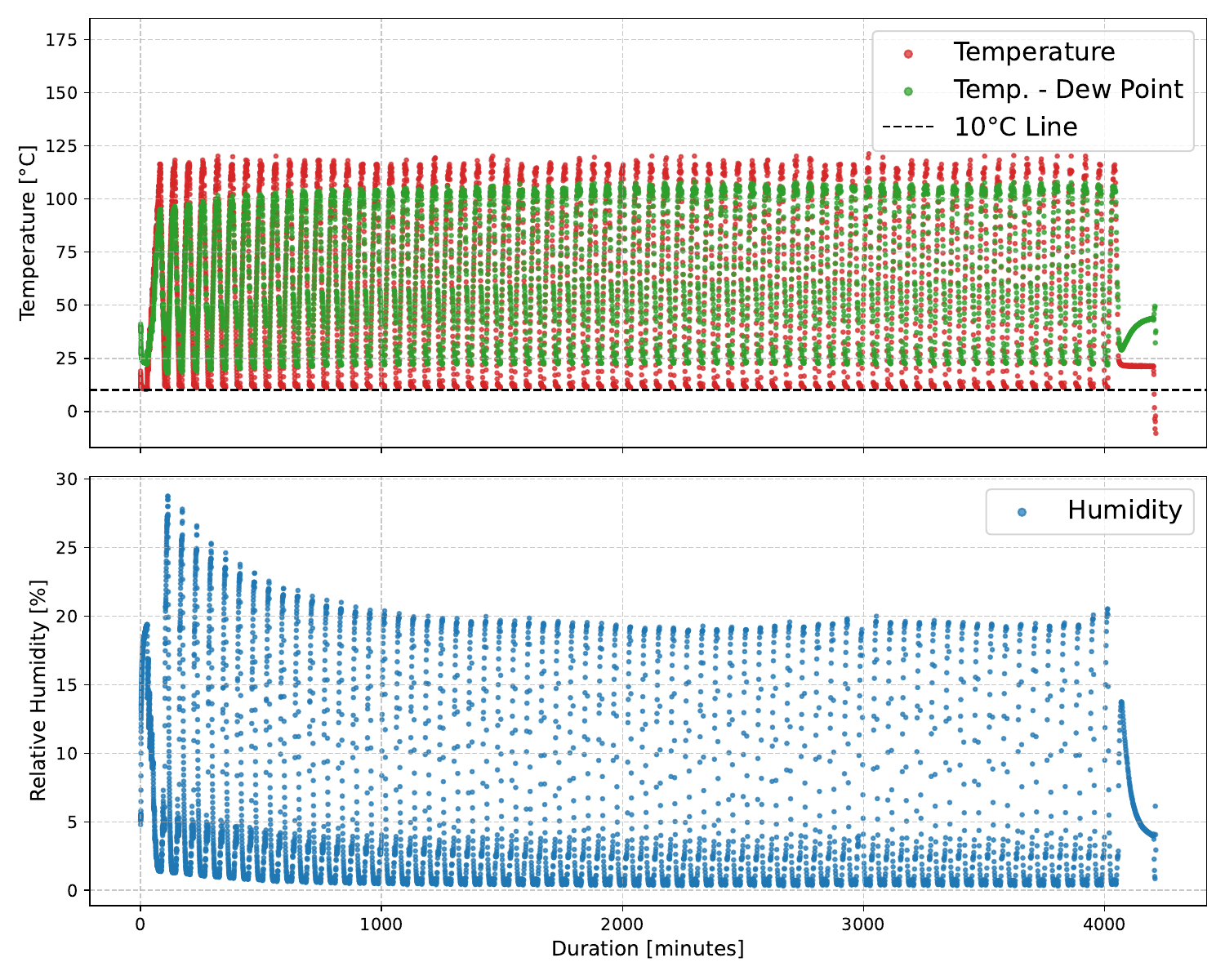}
    \caption{Temperature, dew point and humidity variations as a function of time in a typical Weekend Program. See Fig. \ref{fig:day_program} for the description. In the Weekend Program there was not active N$_2$ flow and the power supply was switched off.}
    \label{fig:weekend_program}
\end{figure}}

In the \textit{Night Program} the silicon sensor was kept at a fixed temperature of +21$^{\circ}$C and at a fixed bias of $-150$~V for an average of 14 hours during weekdays. See Fig.~\ref{fig:overnight_program} for the summary of this program.

In the \textit{Weekend Program} the silicon sensor was thermally shocked by rapidly thermally cycling the climate chamber from +10 to +120 $^{\circ}\text{C}$ for 2 days. The silicon sensor was left unbiased during the weekend. See Fig.~\ref{fig:weekend_program} for the summary of this program.

During all three programs, the humidity in the ambient chamber was controlled such that the dew point was maintained 10 $^{\circ}\text{C}$ below the operating temperature inside the chamber, to avoid condensation. A total of 5 to 8 Day Programs, 7 Night Programs and 1 to 2 Weekend Programs were applied to each sensor. %For the AC-LGAD sensors, the Weekend Programs were limited to 2 and the first day only had one ramp was performed.  If I'm not mistaken there were also only 5 day programs for 3045 and one of those also only had one ramp.
Table~\ref{tab:programs} summarizes the definitions of the Day, Night, and Weekend Programs.

%%%%%%%%%%%%%%%%%%%%%%%%%%%%%%%%%%%%%%%%%%%%%%%%%%%
\subsection{Leakage current versus temperature}
\begin{figure}[t!]
    \centering
    \subfloat[]{\includegraphics[width=.50\linewidth]{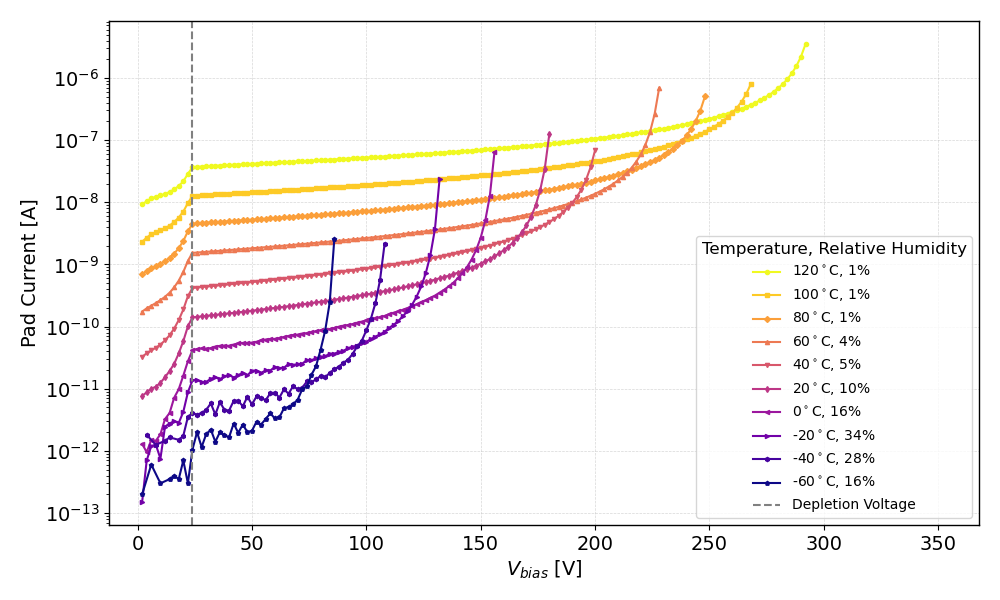}\label{fig:PadI_W3045}}
    \subfloat[]{\includegraphics[width=.50\linewidth]{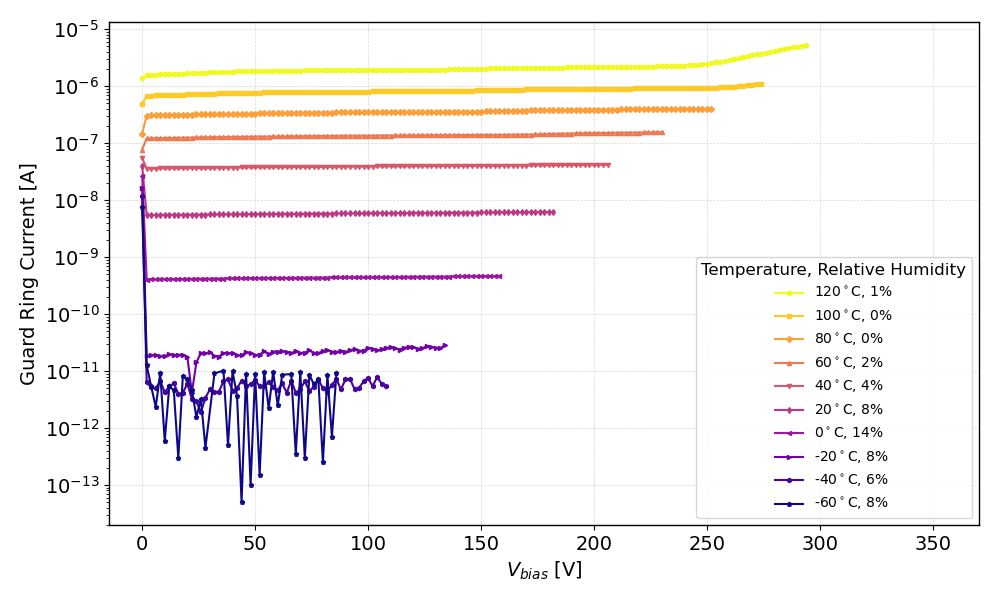}\label{fig:GRI_W3045}}
    \caption{Results of IV scans for pad (a) and guard ring (b) currents for the LGAD W3045 sensor at different temperatures, taken in a Day Program when the temperature was ramped up in steps of 20$ ^{\circ}$C from -60  to +120 $^{\circ}$C. The relative humidity values are reported in the legend.}
    \label{fig:IV_temp_W3045}
\end{figure}

\begin{figure}[t!]
    \centering
    \subfloat[]{\includegraphics[width=.50\linewidth]{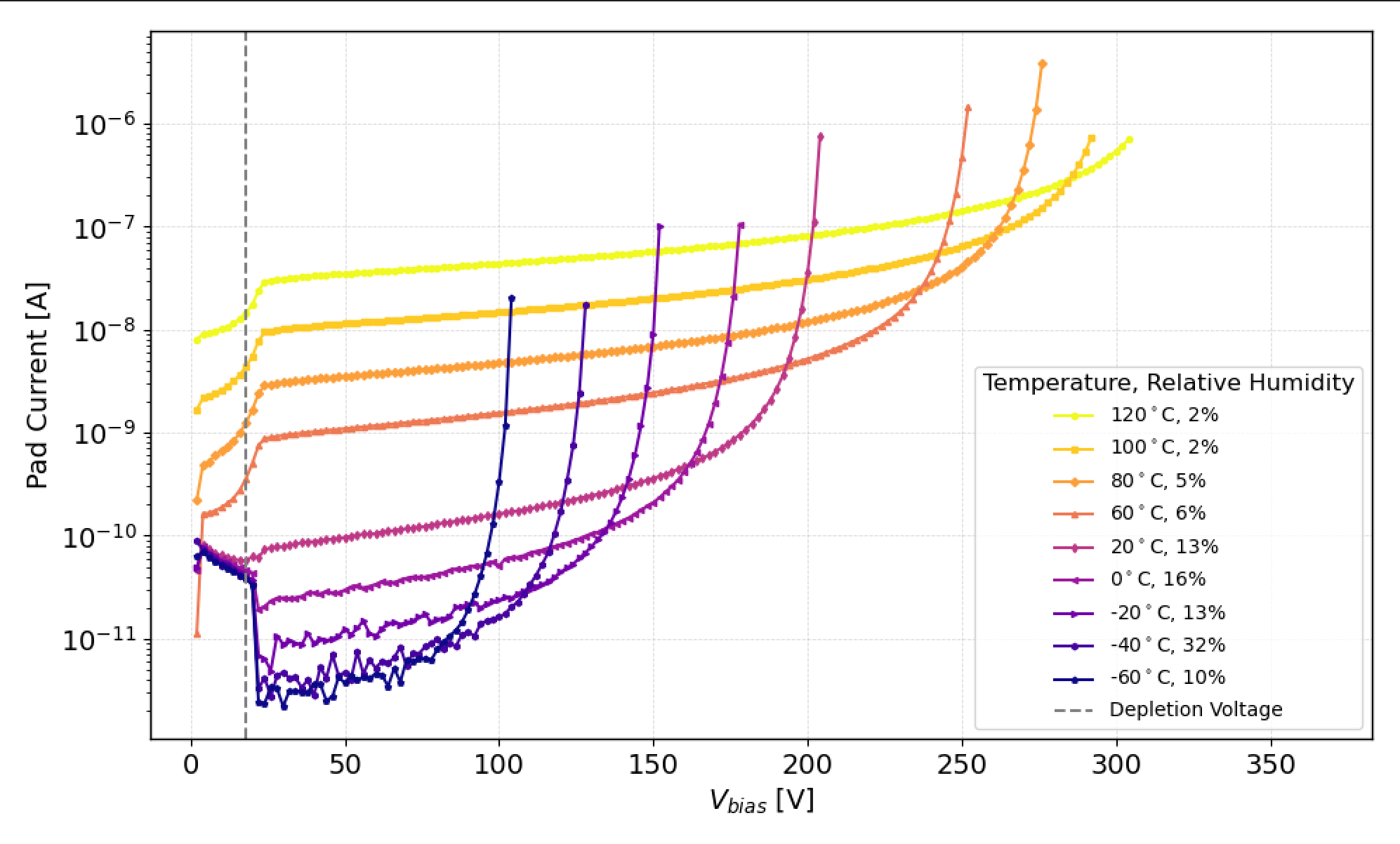}\label{fig:PadI_W3058}}
    \subfloat[]{\includegraphics[width=.50\linewidth]{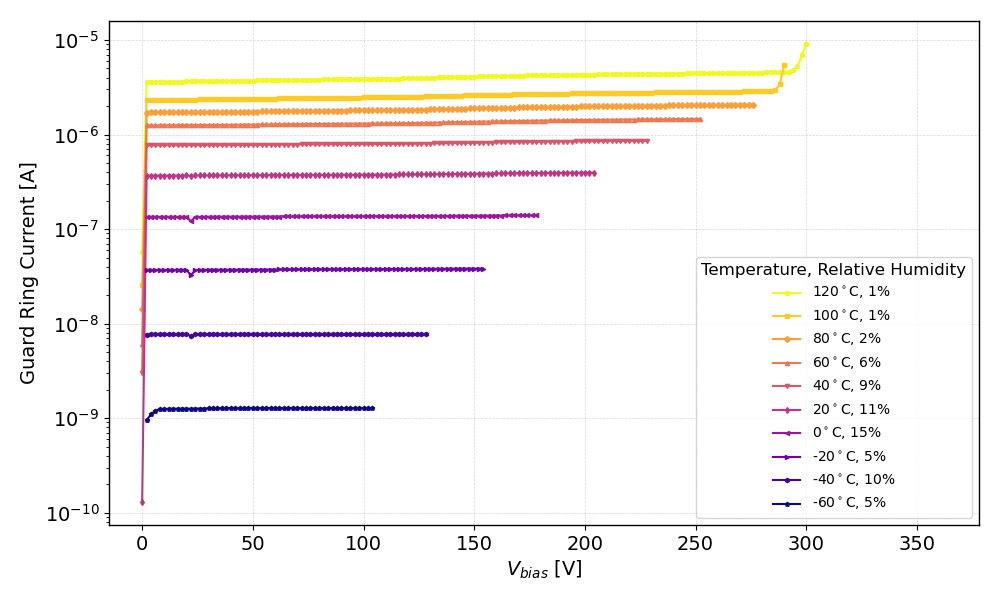}\label{fig:GRI_W3058}}
    \caption{Results of IV scans for pad (a) and guard ring (b) currents for the LGAD W3058 sensor at different temperatures, taken in a Day Program when the temperature was ramped down in steps of 20 $^{\circ}$C from +120  to -60 $^{\circ}$C. The relative humidity values are reported in the legend.}
    \label{fig:IV_temp_W3058}
\end{figure}

%I-V measurements taken during the thermal cycles extends to a wide range of temperatures from $-60 ^0\text{C}$ to $120 ^0\text{C}$.
Figures \ref{fig:PadI_W3045} and \ref{fig:PadI_W3058} show the IV scan results for the pad current of LGAD W3045 and LGAD W3058, respectively, obtained during a Day Program, i.e. the first Day Program as an example, in which the temperature was ramped up in steps of 20 $^{\circ}$C from -60  to +120 $^{\circ}$C (Fig.~\ref{fig:PadI_W3045}) and the temperature was ramped down in steps of 20 $^{\circ}$C from +120  to -60 $^{\circ}$C (Fig.~\ref{fig:PadI_W3058}).
Both the leakage current and the breakdown voltage increase with temperature. No significant changes are observed in the IV scans in the following Day Programs.

The full depletion voltage of the LGAD sensor remains constant at around $-25$~V within the entire temperature range.
Figures \ref{fig:GRI_W3045} and \ref{fig:GRI_W3058} show the IV characteristic of the same sensors for the guard ring current. The guard ring's leakage current monotonically increases with temperature by three to five orders of magnitude in the tested temperature range.
At the two lowest temperature steps, the results show larger fluctuations in the current because the measurement is at the limit of the sensitivity of the readout instrument.
%
%Similar measurements for the pad and guard ring were taken during the temperature ramp-down phase. 
The IV characteristics are the same in ramping up and ramping down.

\begin{figure}[t!]
    \centering
    \subfloat[]{\includegraphics[width=.50\linewidth]{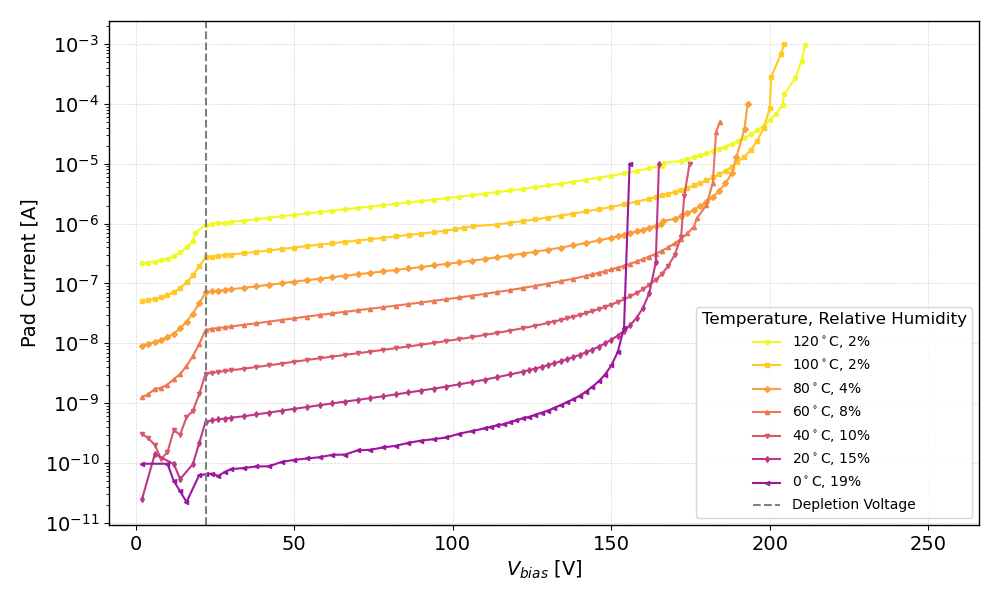}\label{fig:PadI_W3081}}
    \subfloat[]{\includegraphics[width=.50\linewidth]{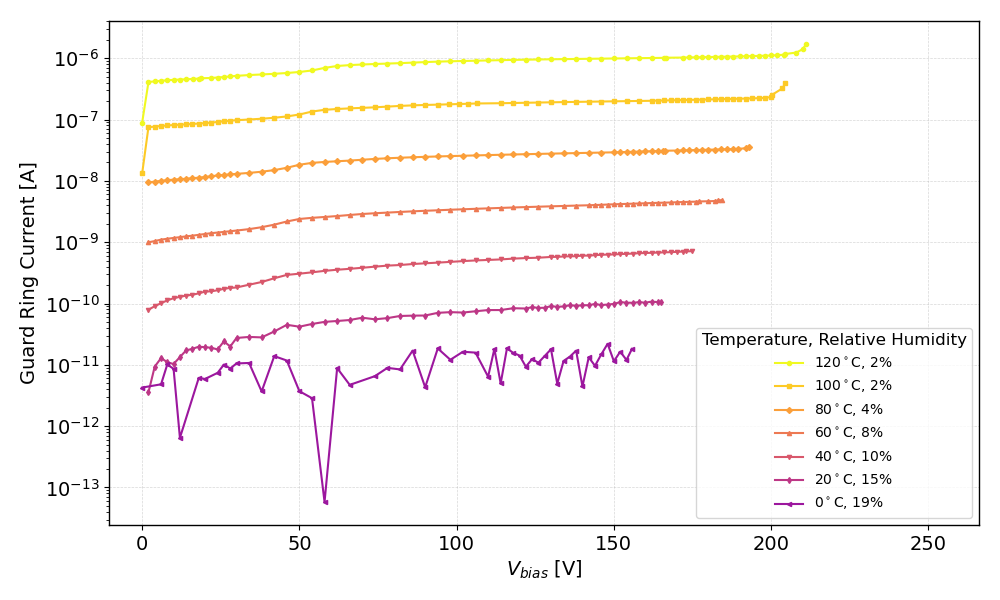}\label{fig:GRI_W3081}}
    \caption{Results of IV scans for pad (a) and guard ring (b) currents for the AC-LGAD W3081 sensor at different temperatures, taken in a Day Program when the temperature was ramped up in steps of 20 $^{\circ}$C from -60  to +120 $^{\circ}$C. Only data at or above 0 $^{\circ}$C are shown due to limited sensitivity of the readout equipment below 100 pA current.}
    \label{fig:AC_IV_temp}
\end{figure}

Figure \ref{fig:AC_IV_temp} shows the IV scan result for the pad and guard ring currents of AC-LGAD W3081 obtained during a Day Program for temperature ramped in steps of 20 $^{\circ}$C from 0 to +120 $^{\circ}$C.
The behavior is similar to that shown in Figs. \ref{fig:IV_temp_W3045} and \ref{fig:IV_temp_W3058} for the LGADs. However, the limited sensitivity of the readout equipment below a current of 100 pA prevents the display of IV curves for temperatures below 0 $^{\circ}$C.
%
%The depletion voltage of the AC-LGAD sensor remains constant at around -25~V within the entire temperature range. Figure \ref{fig:AC_IV_temp} (right) shows the IV characteristic of the same sensor using the guard-ring current I$^{GR}$ during the aforementioned day program. The guard ring leakage current monotonically increases with temperature by five orders of magnitude in the tested temperature range. At the two lowest temperature steps the results show larger fluctuations in the current as the measurement is at the limit of the sensitivity of the readout instrument.
%%%%%%%%%%%%%%%%%%%%%%%%%%%%%%%%%%%%%%%%%%
\subsection{Leakage current versus humidity}
\label{sec:current_vs_humidity}
The electrical response of the LGAD sensors to changing humidity conditions was characterized by varying the humidity inside the climate chamber at a constant temperature. The humidity was changed by controlling the N$_2$ flux and the flow of outside air allowed inside the ambient chamber. This characterization was performed at temperatures between 0 and +25 $^{\circ}$C, in steps of 5 $^{\circ}$C.
The humidity values were recorded at time intervals ranging from 2 to 29 minutes, and the IV measurements were not done in any predefined order.
%
% \begin{figure}[htb]
%     \centering
%     \includegraphics[width=.6\linewidth]{figures/analysis/Humidity_all_IV_at_20.0_pad_Sep13.png}
%     \caption{Results of the IV scan for the pad current (in log$_{10}$ scale), taken at different relative humidity levels for the LGAD W3045 at a constant 20~$^{\circ}$C temperature.}
%     \label{fig:Humidity_vs_current}
% \end{figure}

\begin{figure}[h!]
    \centering
    \includegraphics[width=.49\linewidth]{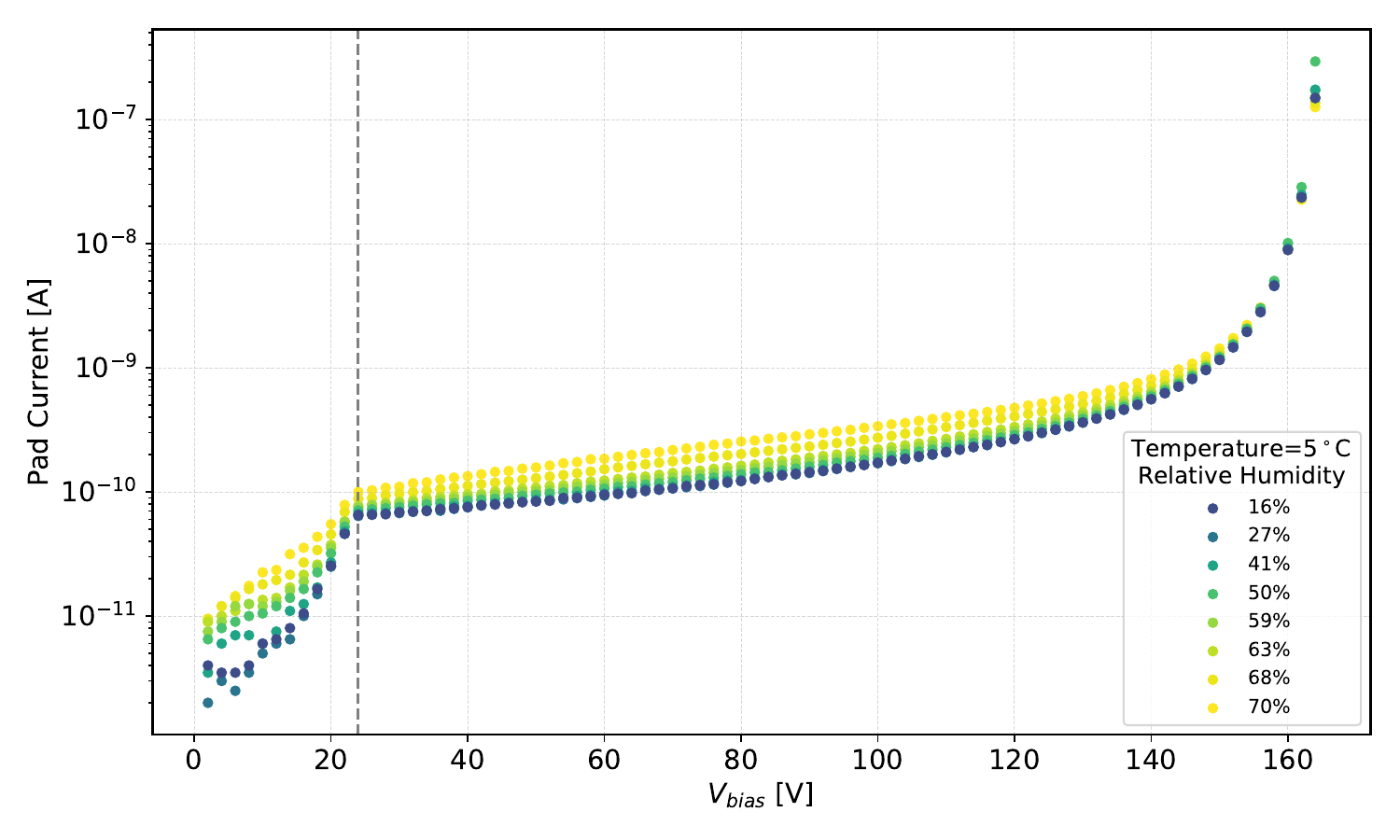}
    \includegraphics[width=.49\linewidth]{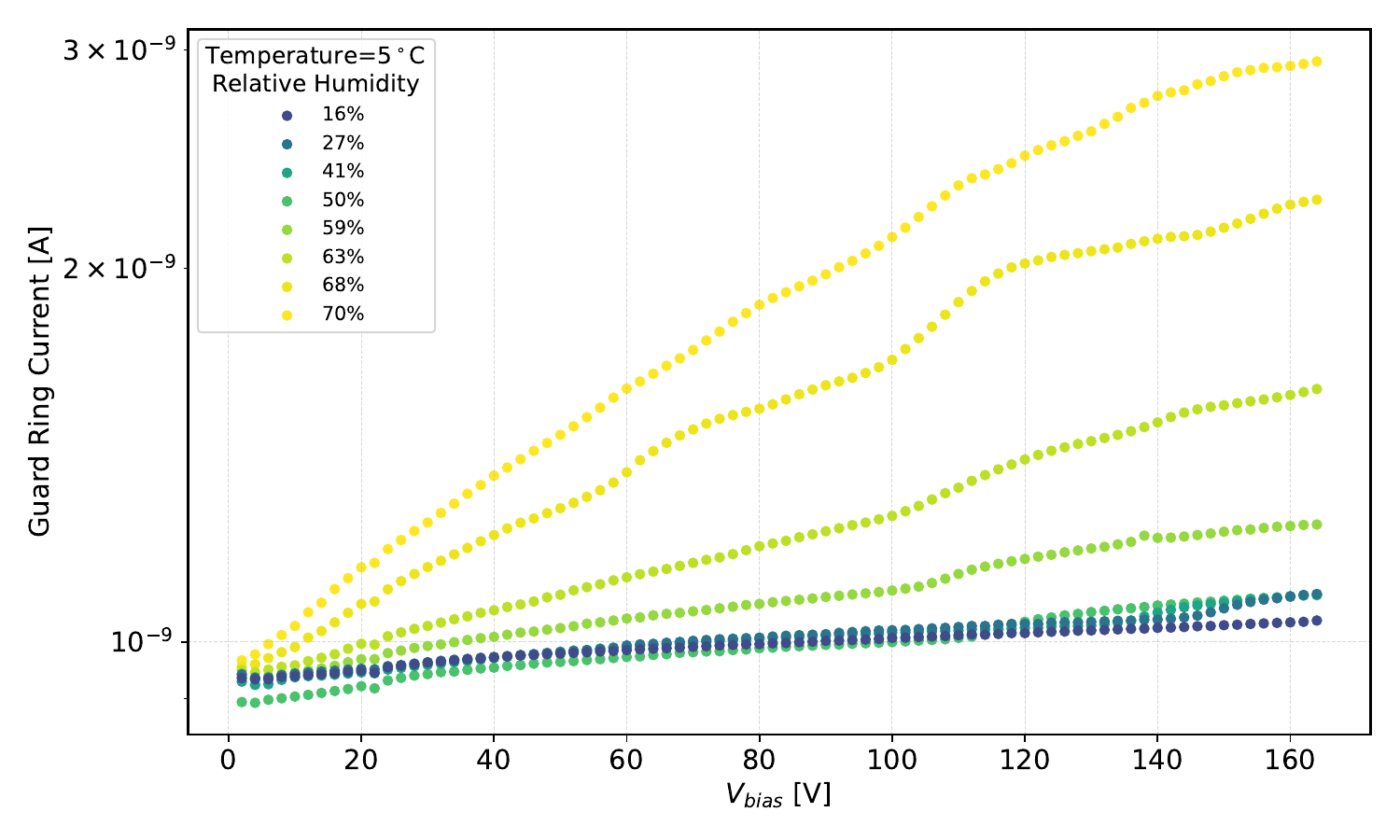}
    \includegraphics[width=.49\linewidth]{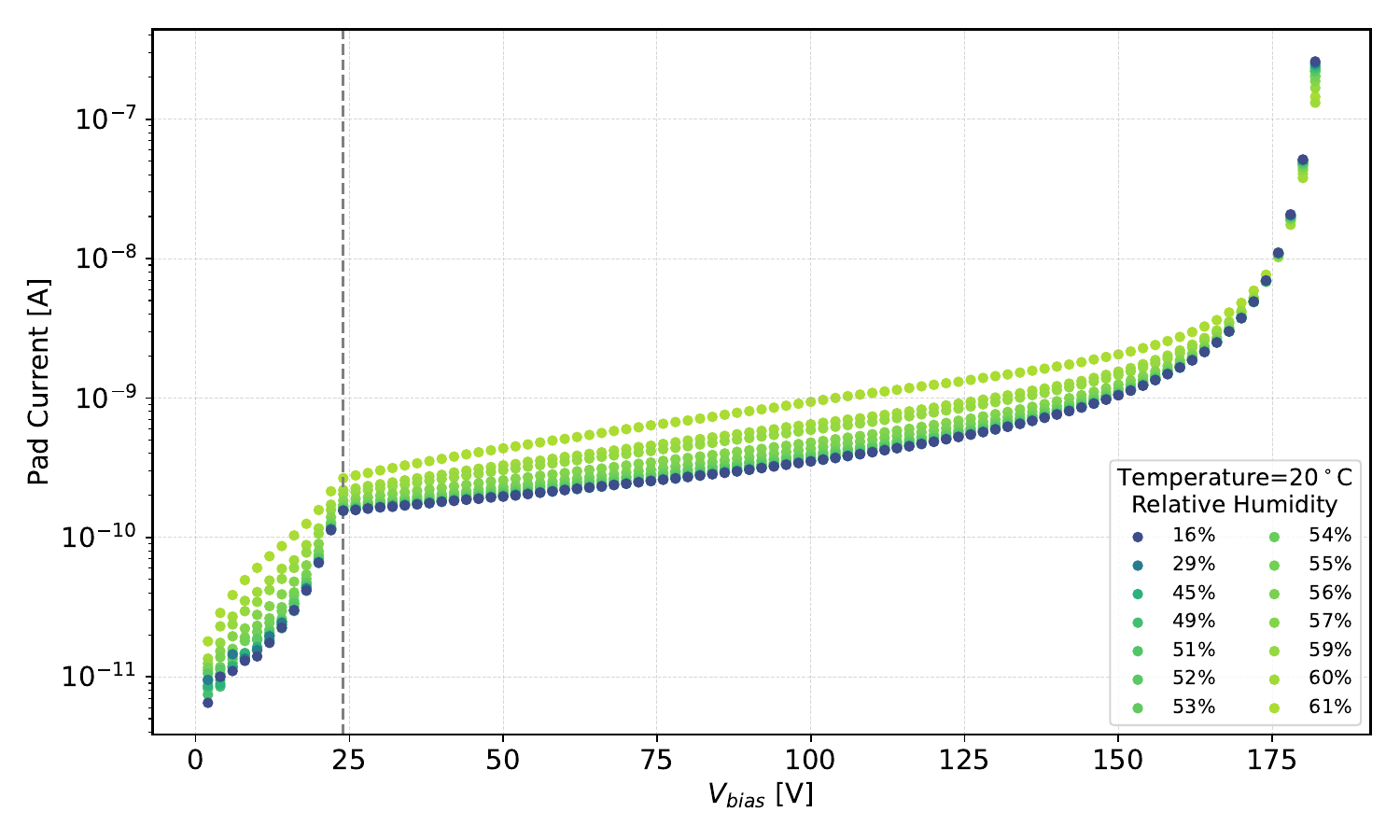}
    \includegraphics[width=.49\linewidth]{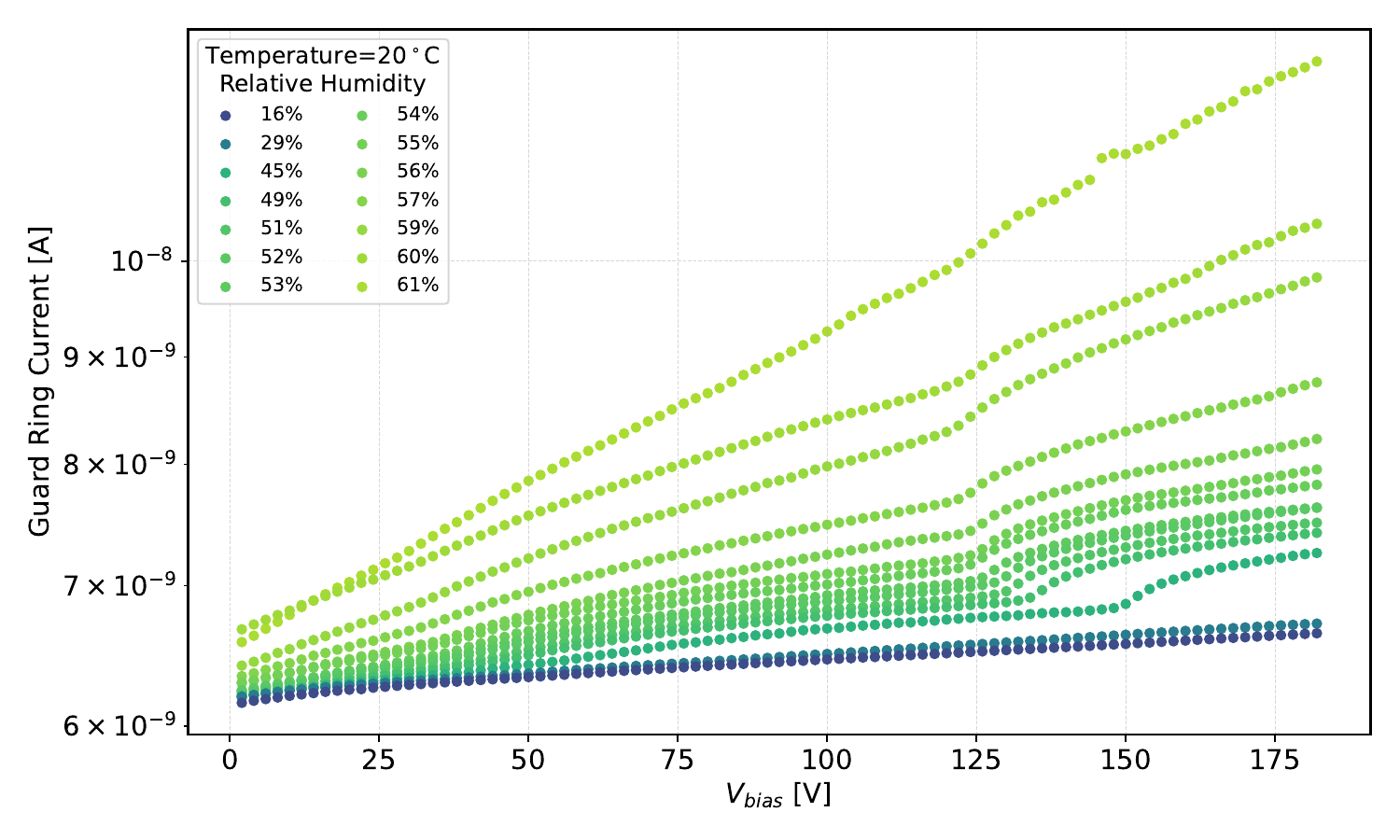}
    \caption{Results of the IV scan for the pad (left) and guard (right) currents, taken at different relative humidity levels for the LGAD W3045, at temperatures 5 (top) and +20 $^{\circ}$C (bottom).}
    \label{fig:Humidity_vs_current}
\end{figure}

Figure~\ref{fig:Humidity_vs_current} shows the pad and guard ring currents for the W3045 LGAD sensor at two representative temperatures, i.e. 5 and 20 $^{\circ}$C, obtained while varying the relative humidity to values up to $70\%$ and $61\%$, respectively. Figure~\ref{fig:Humidity_vs_current} shows that, within the measured range in relative humidity, the breakdown voltage and the full depletion voltage for the pad do not change significantly. However, Fig.~\ref{fig:Humidity_vs_current} shows that the leakage current in the pad plateau region  (between $\sim$25 and $\sim$160~V bias voltage) and in the guard ring increases as a function of humidity for humidity levels higher than about 50\%. The electrical response of the AC-LGAD W3081 sensor to changing humidity was also characterized by varying the relative humidity while the temperature was kept constant at 21 $^{\circ}$C. 
%AT, see Fig.~\ref{fig:AC_Humidity_vs_current}.
%AT: removed following reviewr 'scomment:
%AT \begin{figure}[htb]
%AT    \centering
%AT    \includegraphics[width=.49\linewidth]{figures/analysis/BNL_ACLGAD_W3081_21C_pad_all.png}
%AT    \caption{Results of the IV scan for the pad current (in log$_{10}$ scale), taken at different relative humidity levels for the AC-LGAD at a constant 21 $^{\circ}$C temperature.}
%AT    \label{fig:AC_Humidity_vs_current}
%AT\end{figure}
%
 The relative humidity was varied over a smaller range than for the LGAD --- between 0 and 36\% --- due to low atmospheric humidity when the test was carried out. Similarly to the results for the LGAD, the leakage current, the breakdown and depletion voltages do not change significantly with humidity, within the measured range, i.e. with the maximum humidity lower than the 50\% point at which the leakage current starts to increase for the LGAD sensor.
%%%%%%%%%%%%%%%%%%%%%%%%%%%%%%%%%%%%%%%%%%%%%%%%%%%%%%
\section{Experimental results}\label{sec:analysis}
\begin{figure}[t!]
    \centering
    \includegraphics[width=.48\linewidth]{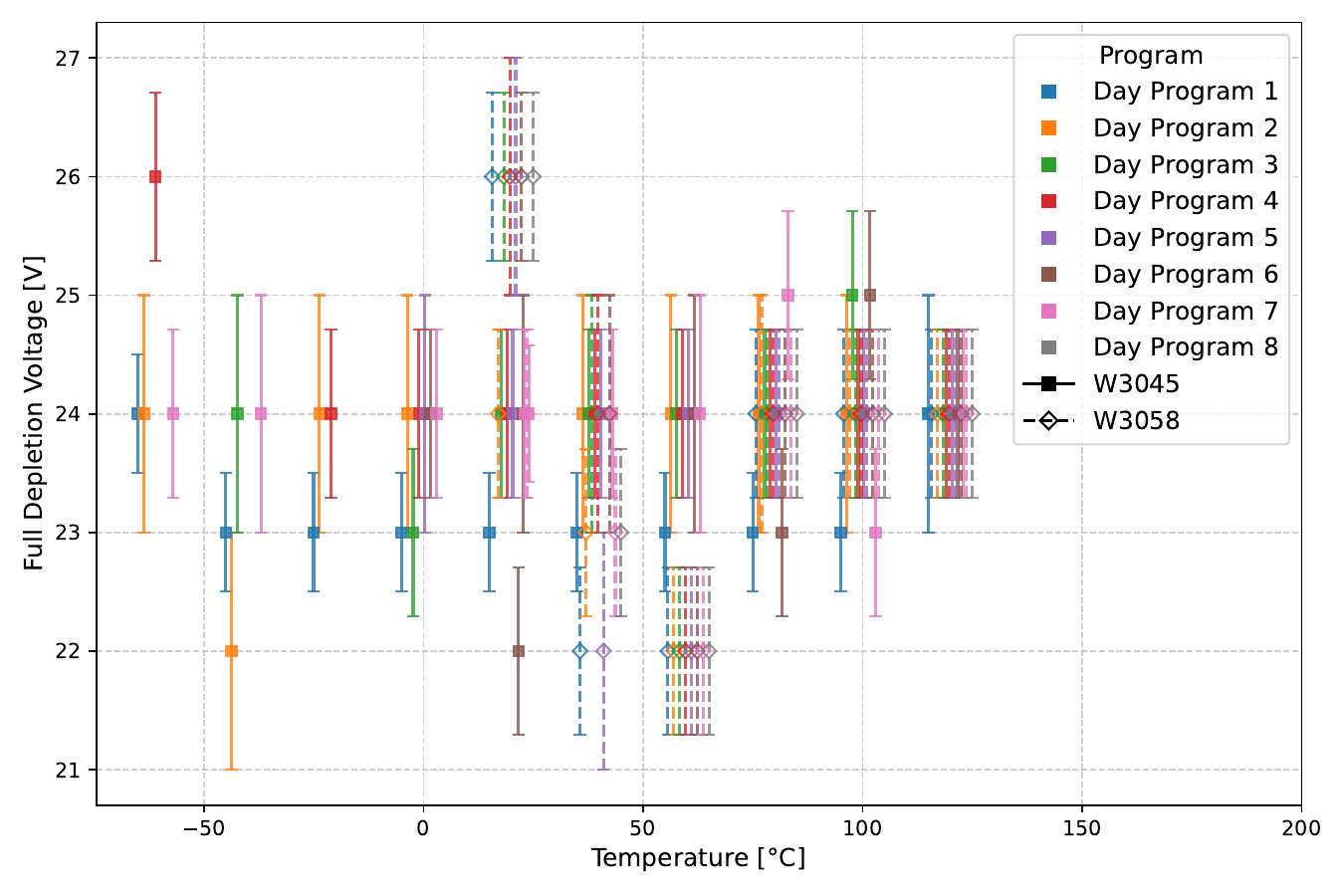}
    \caption{%Breakdown voltage (left)
    Full depletion voltage as a function of temperature across all thermal cycles for each LGAD sensor. Some of the data points are shifted along the temperature axis for visibility purposes.}
    \label{fig:temp_vs_depletion}
\end{figure}

%
%As seen in Fig \ref{fig:IV_temp_pad} and Fig \ref{fig:Humidity_vs_current} the depletion voltage of the sensor is independent of temperature and humidity.
The full depletion voltage is calculated by estimating the second derivative of an IV measurement and averaging the result over the temperature ramp-up and ramp-down  data sets. The uncertainty is estimated by propagating an uncertainty of half of a voltage step through the average calculation. The full depletion voltage is found to be independent of temperature and humidity, and to remain constant in different cycles, for all tested silicon sensors, as is shown in Figs.~\ref{fig:Humidity_vs_current} %,~\ref{fig:AC_Humidity_vs_current} 
 and~\ref{fig:temp_vs_depletion}. The full depletion voltage depends only on the fabrication parameters, for example doping concentrations, of each sensor.

\begin{figure}[t!]
    \centering
    \includegraphics[width=.8\linewidth]{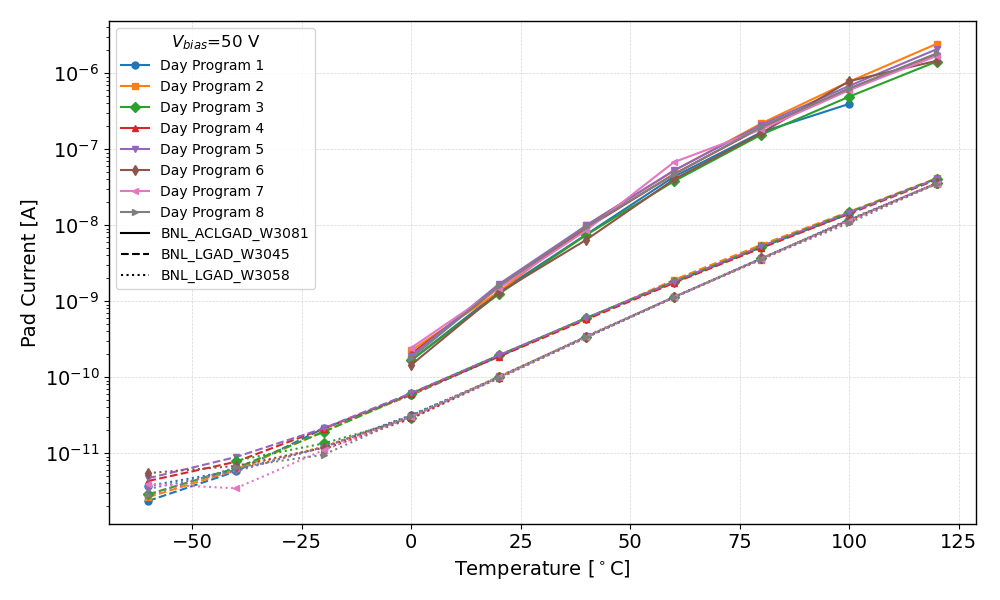}
    \caption{Pad leakage current as a function of temperature at the operating voltage of 50 V for two LGADs and one AC-LGAD in various Day Programs.}
    \label{fig:leakage_vs_temp_50V}
\end{figure}

Conversely, the pad current increases exponentially with temperature for the LGADs and AC-LGAD, as is shown in Fig.~\ref{fig:leakage_vs_temp_50V}. In this study, the pad current is displayed for an operating bias voltage fixed at 50 V; that is on the IV curve plateau for all temperatures.
All sensors show consistent leakage current over multiple cycles.  The LGAD from W3058 wafer shows lower leakage current compared to that from the W3045 wafer. This is a consequence of the improved design of the W3058 wafer, for instance the LGAD from W3045 wafer lacks the $p^+$ termination at the scribeline that stops the surface electron layer and the surface depletion from reaching the cut edge. The AC-LGAD from wafer W3081 shows higher leakage current at high temperature and a steeper temperature dependence overall compared to the LGADs. These effects are due to slightly different fabrication processes between AC-LGADs and LGADs.

\begin{figure}[t!]
    \centering
    \includegraphics[width=.8\linewidth]{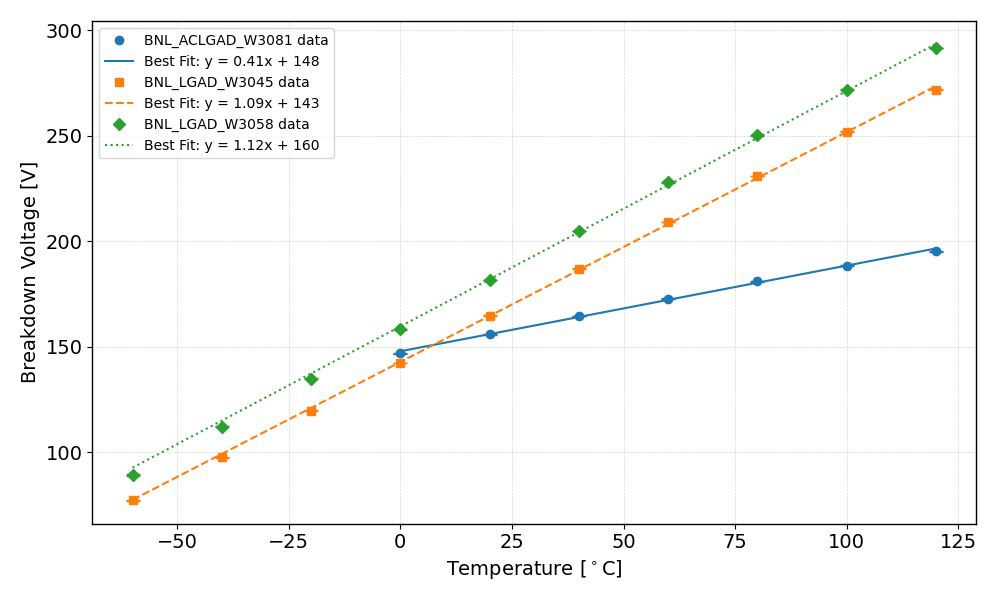}
    \caption{Breakdown voltage as a function of temperature, averaged over all thermal cycles for each sensor.}
    \label{fig:temp_vs_breakdown}
\end{figure}

\begin{figure}[htb!]
    \centering
    \includegraphics[width=.6\linewidth]{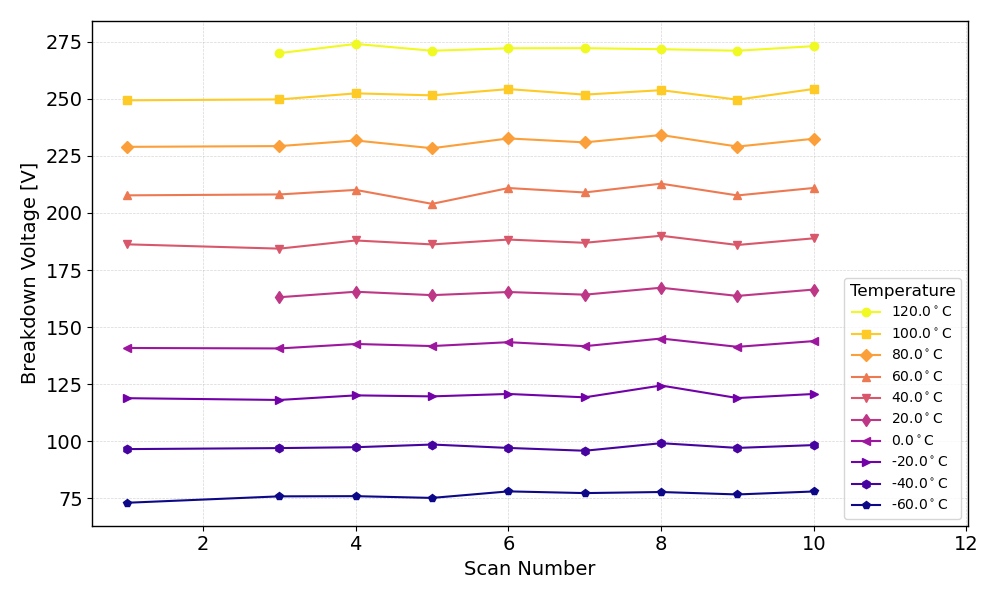}
    \includegraphics[width=.6\linewidth]{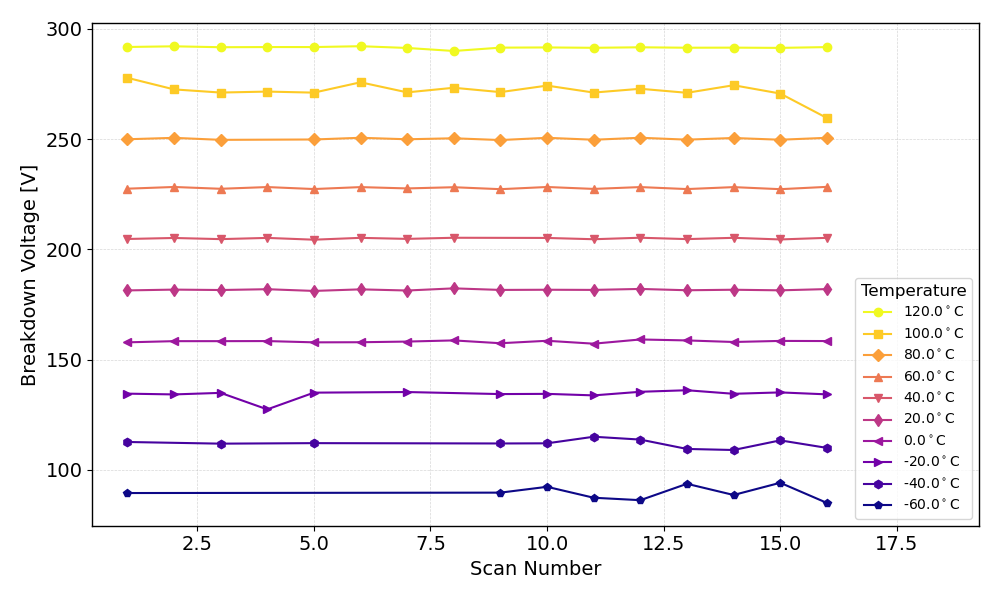}
    \includegraphics[width=.6\linewidth]{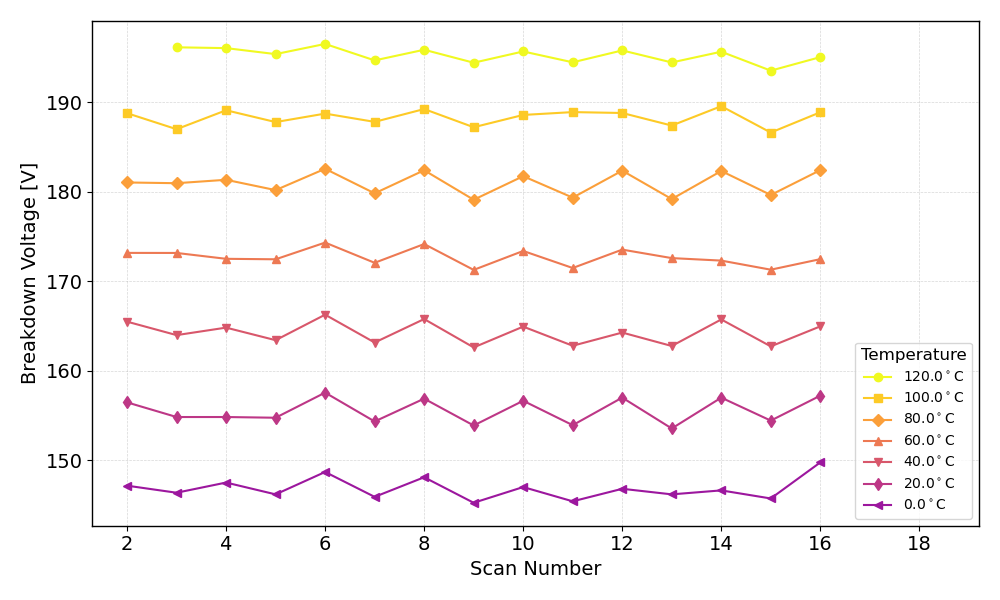}
    \caption{Breakdown voltage as a function of the Day Program thermal cycle number for LGAD sensors W3045 (top) and W3058 (center) and AC-LGAD W3081 (bottom) at different temperatures.}
    \label{fig:breakdown_trend}
\end{figure}

Traditionally, the breakdown voltage (V$_{\mathrm{BD}}$) is estimated by fitting lines in the plateau region and in the post-breakdown region and denoting V$_{\mathrm{BD}}$ as the voltage at which these intersect. For limited post-breakdown data, this would result in very high uncertainties. 
Instead, the breakdown voltage is estimated using a novel method described in Appendix~\ref{app:vBD-method}.
The breakdown voltage increases linearly with temperature, and no dependence on the number of thermal cycles is observed; this is shown in Figs.~\ref{fig:temp_vs_breakdown} %~\ref{fig:temp_vs_breakdown_cycles} 
 and ~\ref{fig:breakdown_trend}. The slope of the linear function of breakdown voltage as a function of temperature is similar between the two LGADs ($1.09\pm 0.01$ and $1.12\pm 0.01$ V/$^{\circ}$C for W3045 and 3058, respectively).  That of the AC-LGAD is $0.41\pm 0.01$~V/$^{\circ}$C, lower than those of the two LGADs. It is observed that the difference in slopes between the AC-LGAD and the LGADs scales with the sensor thicknesses: 20 $\mu\textrm{m}$ / 50 $\mu\textrm{m}$. 

%This is potentially due to other differences between the sensors such as thickness and annealing time.

%\textcolor{red}{AT: add here normalized slopes, as done by Timo in simulation.}\\

%\textcolor{red}{AT: add here a paragraph on the measurements done at CERN  (as a subsection)? add a table with combibed results?}\

% \begin{figure}[htbp]

%
% \begin{figure}[htbp]
%\begin{figure}[htb!]
%\    \centering
%\    \includegraphics[width=.6\linewidth]{figures/analysis/W3045_all_bd.png}
%\    \includegraphics[width=.6\linewidth]{figures/analysis/W3058_all_bd.png}
%\    \includegraphics[width=.6\linewidth]{figures/analysis/W3081_all_bd.png}
%\    \caption{Breakdown voltage as a function of temperature after several Day Program thermal cycles for LGAD sensors W3045 (top) and W3058 (middle) and AC-LGAD W3081 (bottom).}
%\    \label{fig:temp_vs_breakdown_cycles}
%\\end{figure}

% \begin{figure}[htbp]

\begin{figure}[ht!]
    \centering
    \includegraphics[width=.45\linewidth]{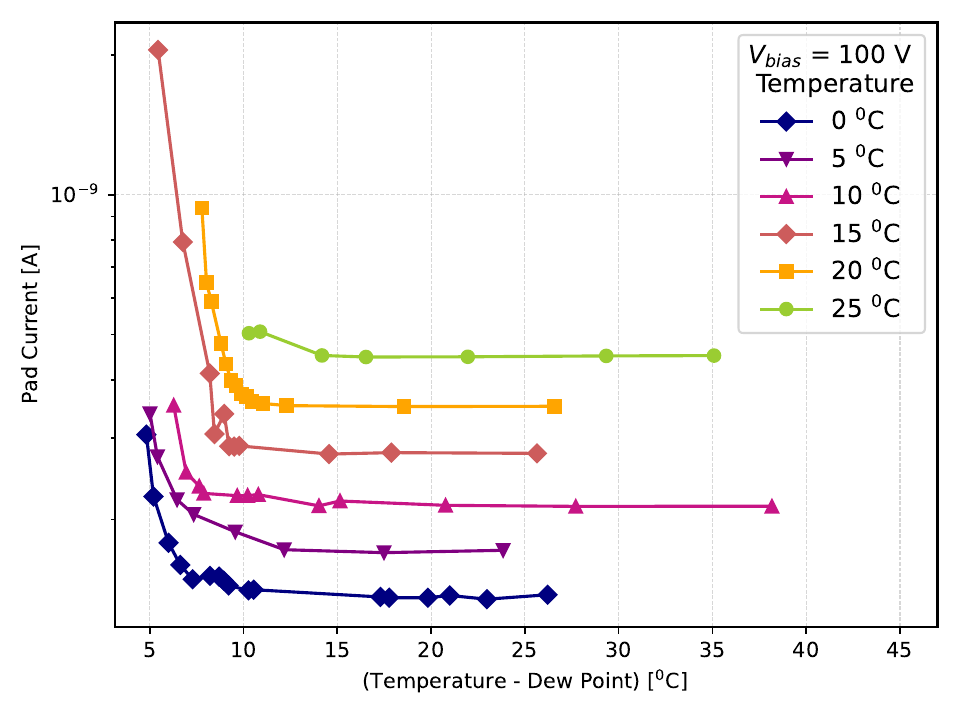}
    \includegraphics[width=.45\linewidth]{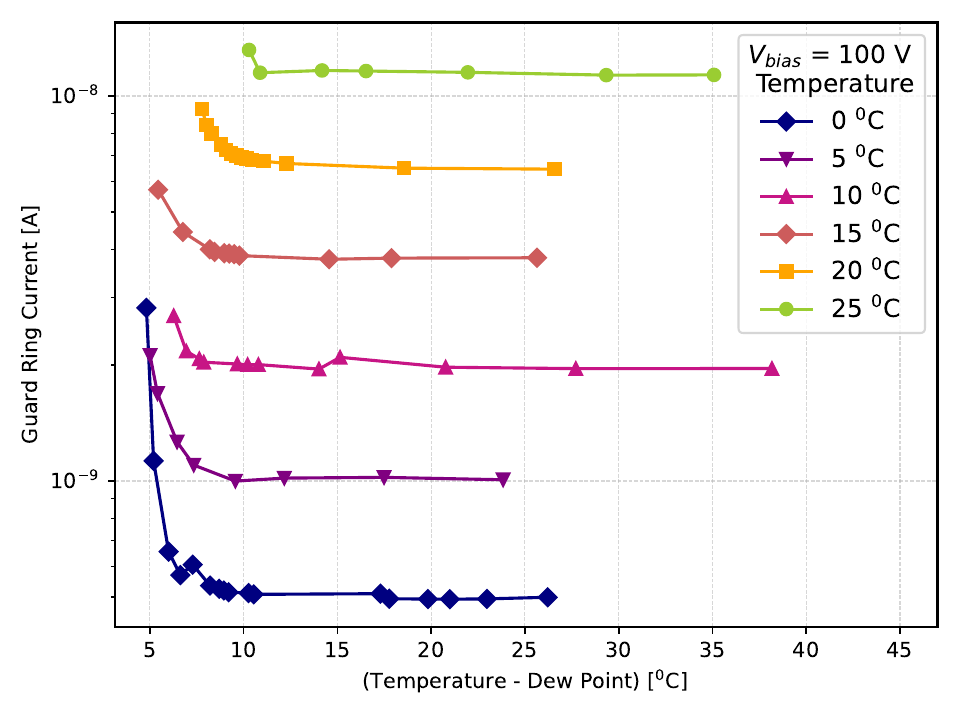}
    \caption{Leakage current of the pad (left) and guard ring (right) for the LGAD W3045 sensor measured as a function of the difference between the temperature and the dew point, when the sensor is biased at 100 V. The results for the same sensor biased at 50 V are not shown as they are qualitatively the same.}
    \label{fig:current_vs_dewpoint1}
\end{figure}

\begin{figure}[ht!]
    \centering
    \includegraphics[width=.45\linewidth]{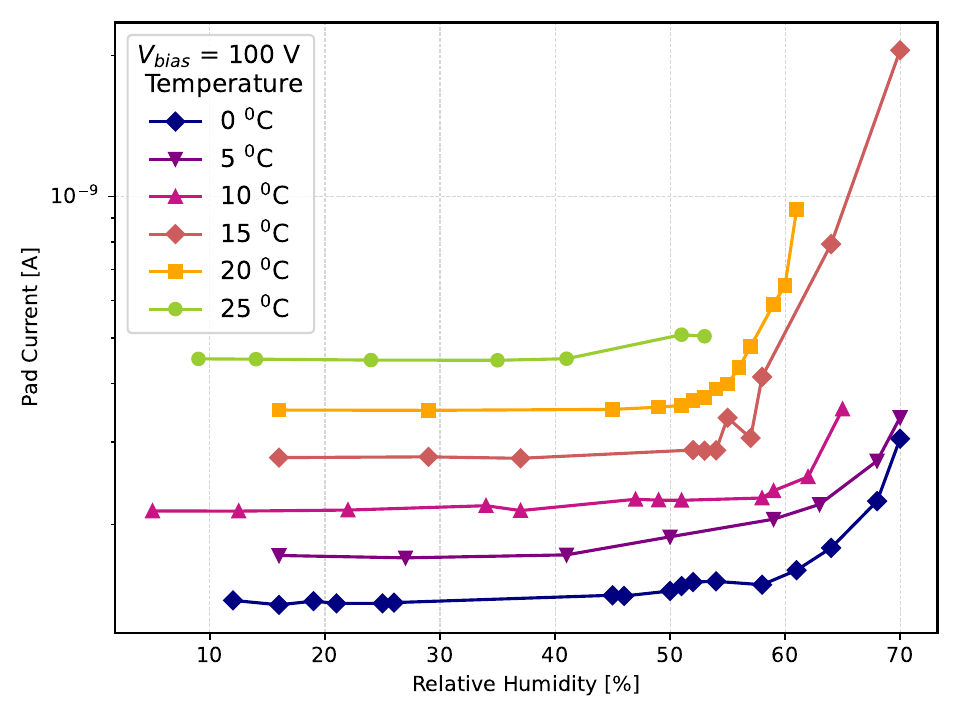}
    \includegraphics[width=.45\linewidth]{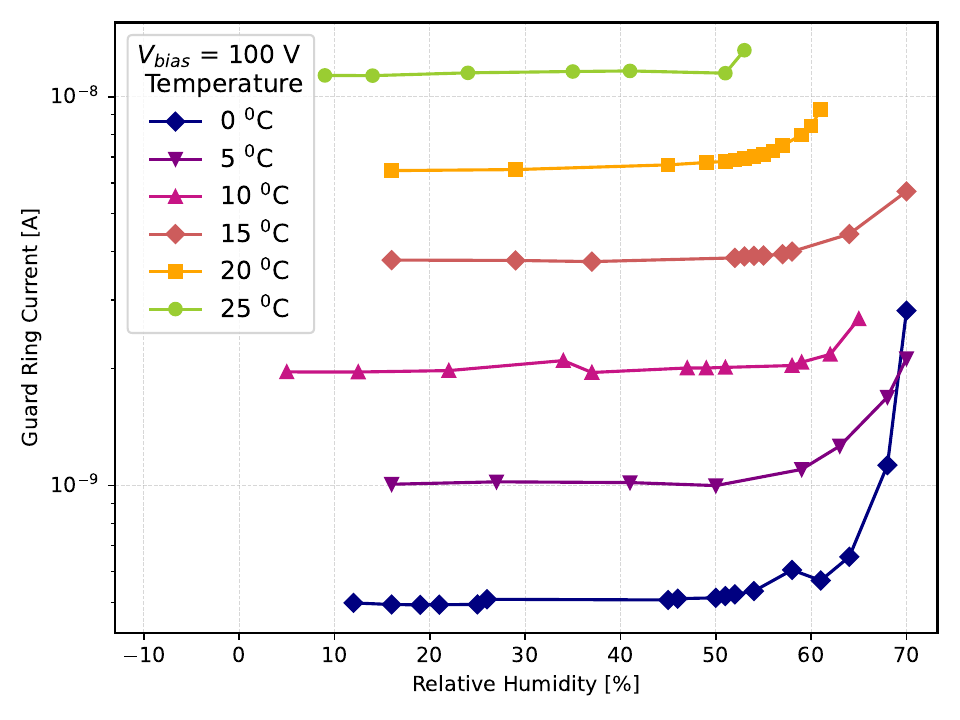}
    \caption{Leakage current of the pad (left) and guard ring (right) for the LGAD W3045 sensor measured as a function of humidity, when the sensor is biased at 100 V. The results for the same sensor biased at 50 V are not shown as they are qualitatively the same.}
    \label{fig:current_vs_dewpoint2}
\end{figure}

To quantify the effect of humidity on the silicon sensor currents, the pad and guard ring leakage currents are studied as a function of humidity at fixed values of operating bias voltage, 50 V and 100 V, for temperatures in the range 0 to +20 $^{\circ}$C in steps of 5 $^{\circ}$C. For each data point, the dew point was also calculated. 
Figure~\ref{fig:current_vs_dewpoint1} shows the pad and guard ring leakage currents as a function of the difference between the temperature and the dew point, which indicates the proximity to condensation. For both pad and guard ring, when the temperature is less than about 10 $^{\circ}$C above the dew point, the leakage current rapidly increases. Since the relative humidity is determined by the ratio of the saturation vapour pressures at the dew point and at the sensor temperature, a 10 $^{\circ}$C difference corresponds to a relative humidity of about 47–54\% over the 0–25 $^{\circ}$C range studied~\cite{sensirion2011sht7x}, as illustrated in Fig.~\ref{fig:current_vs_dewpoint2}, which shows the pad and guard ring currents as a function of the relative humidity. The onset observed in Figs.~\ref{fig:current_vs_dewpoint1} and~\ref{fig:current_vs_dewpoint2} is therefore the same condition as the approx. 50\% relative humidity threshold found in Sec.~\ref{sec:current_vs_humidity}. This is consistent with the surface effects discussed in Sec.~\ref{sec:theory_humidity}, which are governed by the amount of water adsorbed on the oxide surface and hence by the relative humidity. We note that the 10 $^{\circ}$C dew-point margin adopted during the thermal-cycling programs (Sec.~\ref{sec:thremal_cycling}) keeps the sensors just below this threshold.

Comparing the performance of the two LGADs and one AC-LGAD tested, over the course of multiple Day, Night, and Weekend Programs, the values of depletion voltage, leakage current, and breakdown voltage remain consistent.
%%%%%%%%%%%%%%%%%%%%%%%%%%%%%%%%%%%%%%%%%%%%%%%%%%
\section{Simulations and Data Interpretation}\label{sec:simulations}
 Two-dimensional simulations were carried out by using the Synopsys Sentaurus\footnote{http://www.synopsys.com} finite-element Technology Computer-Aided Design (TCAD) software framework.
 Since an accurate model of impact ionization in the sensor gain layer is critical for the modeling of any device with an internal avalanche mechanism, a survey study of several impact ionization models (van Overstraeten-de Man \cite{vO1970}, Okuto-Crowell \cite{Okuto1975}, Lackner \cite{Lackner1991}, Bologna, New Bologna \cite{UniBo2005,UniBo2_2004} and Massey~\cite{Massey2006}) was carried out before the simulations presented here were concluded.
 The following simulation results were produced with the Massey model, % \cite{Massey2006}, 
 which displayed the closest agreement with the measured temperature % dependence of the avalanche onset.of the temperature 
effects on the avalanche onset \cite{Peltola2026_DRD3}.
%
% \textcolor{red}{AT: expand on this model survey? e.g. show a plot that shows that Massey model is the best? Or add a reference?}
%
%%%%%%
\subsection{Simulation device structure and parameters}
\begin{figure}[h!]
\centering
% \subfloat[]{\includegraphics[width=3.0in]{pSample.png}\label{pSample}}\\%
% \hfil
% \subfloat[]{\includegraphics[width=.54\textwidth]{Rint_device2.png}\label{interPad}}\hspace{1mm}%
% \subfloat[]{\includegraphics[width=.45\textwidth]{HGCALdoping3a.png}\label{HGCALdoping}}%
\subfloat[]{\includegraphics[width=.60\textwidth]{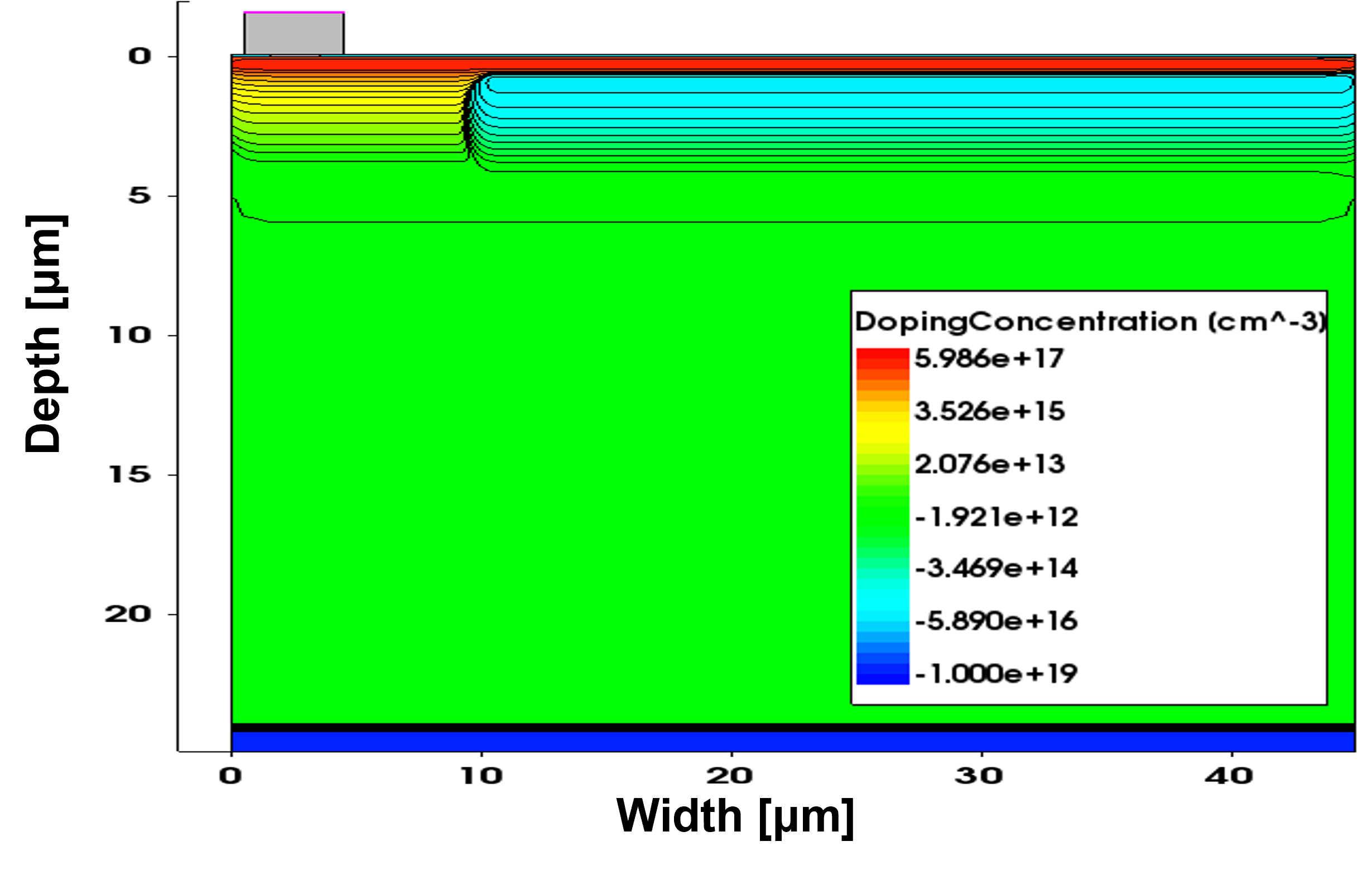} \label{LGAD_2D}}
%\hspace{1mm}%
\\
%% \subfloat[]{\includegraphics[width=.49\textwidth]{LGAD_BNL_2Db.png}\label{LGAD_2D}}% \hspace{1mm}%
% \subfloat[]{\includegraphics[width=.51\textwidth]{figures/Doping_20_50um_BNL2.png}\label{LGADdoping}}%
\subfloat[]{\includegraphics[width=.5\textwidth]{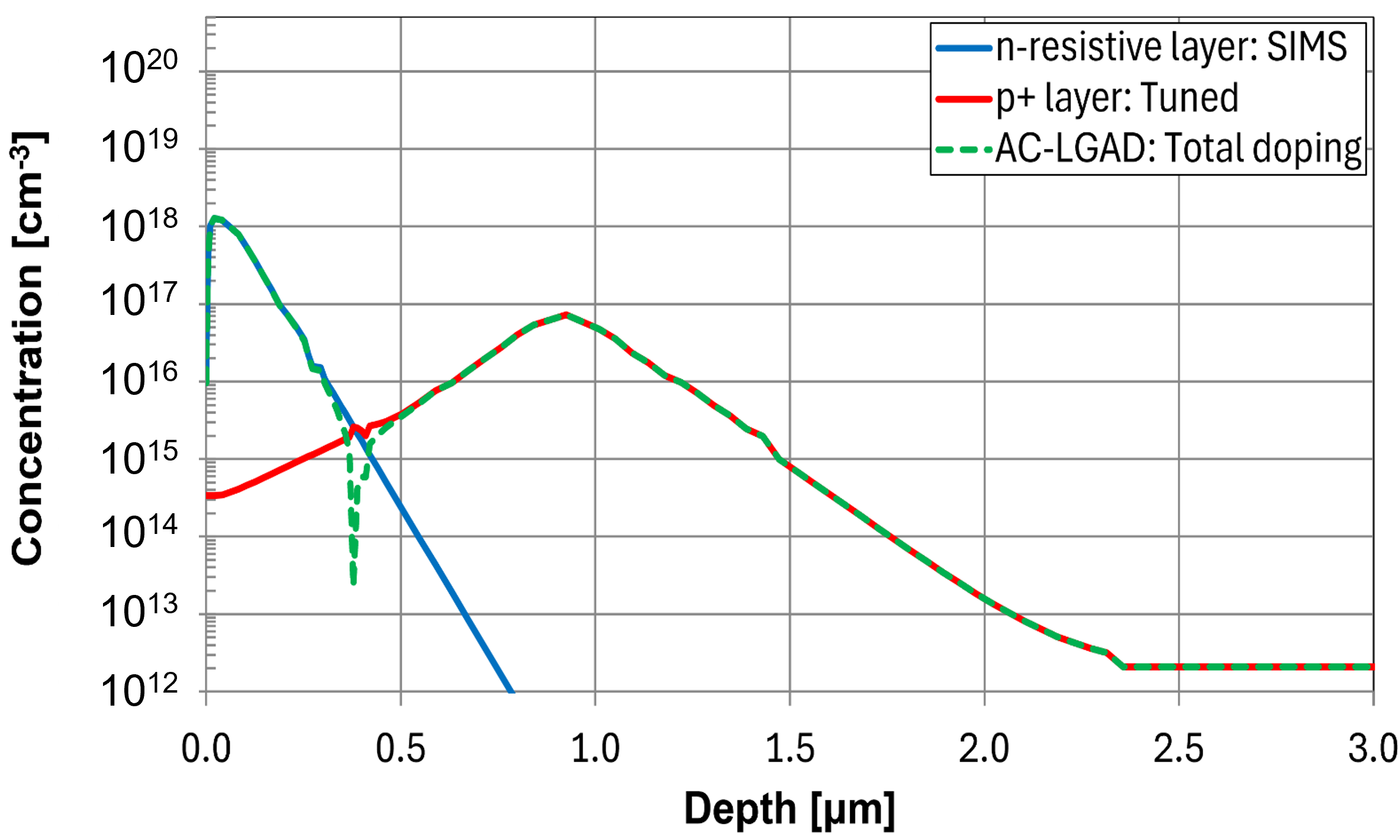}\label{LGADdoping}}%
\subfloat[]{\includegraphics[width=.5\textwidth]{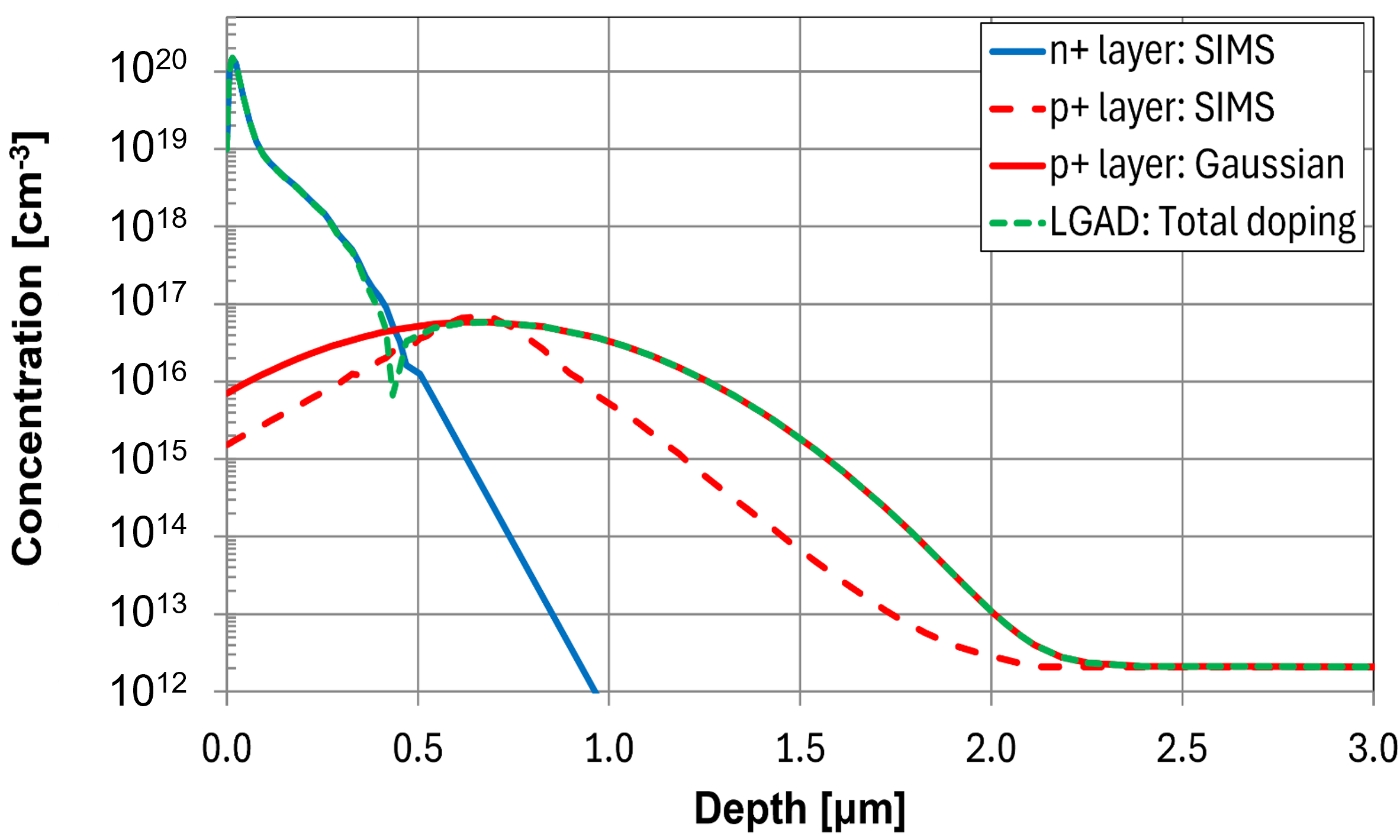}\label{LGADdoping50}}%
\caption{\small (a) % Active area of the 
Modeled edge region and doping concentration of an AC-LGAD with $p$-type bulk of $20~\mu\textrm{m}$ thickness. The backplane $p^{++}$-substrate and aluminum contact are not pictured. (b) SIMS-measured $n$-resistive and simulation-tuned $p^+$ layers % from BNL wafers W3051 and W3049, respectively, 
and the resulting TCAD-input total doping profile at the gain layer of an AC-LGAD with a $20~\mu\textrm{m}$-thick epi-layer. 
% (c) Corresponding doping profiles for $t_\textrm{epi}=50~\upmu\textrm{m}$ for both AC-LGAD and LGAD-like peak doping levels at the $n$-layer. 
(c) SIMS-measured $n^+$ and $p^+$ layers % from BNL wafers W3076 and W3051, respectively, 
and an analytical $p^+$ layer with Gaussian decay for an LGAD with a $50~\mu\textrm{m}$-thick epi-layer. The % resulting TCAD input 
total doping profile combines the SIMS $n^+$ and the analytical $p^+$ layers.
}
\label{LGAD_TCAD_doping}
\end{figure}

% The 2D device-simulations % presented in this paper 
% were carried out using the Synopsys Sentaurus\footnote{http://www.synopsys.com} finite-element Technology Computer-Aided Design (TCAD) software framework. Massey \cite{Massey2006}...
%
To model % and investigate
the temperature dependence of the pad leakage currents and breakdown voltages of the LGAD and AC-LGAD in Figs.~\ref{fig:PadI_W3058} and~\ref{fig:PadI_W3081}, respectively,
% For the 
% a generic AC-LGAD's % simulations the 
% edge-region AC-
a sensor structure displayed in Figure~\ref{LGAD_2D} % and the inter-cell region in Figure 3a were 
was implemented. Since only the DC-coupled biasing electrode (with 1.5~$\mu\textrm{m}$ thick aluminum) % \textcolor{red}{(AT: define $t_{al}$)} 
outside the gain layer was implemented on the front surface, the same structure was used to model both the 50~$\mu\textrm{m}$-thick LGAD and the 20~$\mu\textrm{m}$-thick AC-LGAD. The electrode was connected to the epi-layer % by a via 
through a uniform 100~nm-thick SiO$_2$ layer. The high potential of the reverse bias voltage was supplied from the backplane with a 1.5 and 20~$\mu\textrm{m}$-thick uniform aluminum-contact and low-resistivity $p^{++}$ substrate, respectively. % \textcolor{red}{AT: what does 'respectively refer to? LGAD vs AC-LGAD?}.

The doping profiles at the gain layer ($n$-resistive and $n^+$ layer for AC-LGAD and LGAD, respectively, and $p^+$-layer) and the resulting total doping concentrations for both epitaxial thicknesses are shown in Figs.~\ref{LGADdoping} and~\ref{LGADdoping50}. The doping profiles of the $n$-resistive and $n^+$ layers were implemented into the simulation directly from the Secondary Ion Mass Spectrometry (SIMS) data of BNL-produced sensors. % \textcolor{red}{Also shown is the Secondary Ion Mass Spectrometry (SIMS) measured doping profile\footnote{Typical dynamic range of SIMS is 3--4 decades.} of a BNL-produced AC-LGAD's $n$ resistive layer, generated by phosphorus-ion implantation with energy and dose of 40 keV and $2\times10^{13}~\textrm{cm}^{-2}$, respectively. Displayed in Figs.~\ref{LGADdoping} and~\ref{LGADdoping50}, the $n$ resistive layer doping profile was reproduced by Sentaurus TCAD process simulation and implemented for both thickness sensors. Since the measured 50-$\mu\textrm{m}$ device was LGAD with $n^+$ layer instead of $n$-resistive layer, a second gain-layer doping profile for this thickness was also implemented in Figure~\ref{LGADdoping50}, with the analytical $n^+$ layer following Gaussian decay.} 
%The final doping profiles for each thickness were reached by tuning the $p^+$-layer peak doping value ($N_\textrm{p}$) and depth, and the epitaxial layer doping value, to reproduce
%% full depletion voltage ($V_\textrm{fd}$) and 
%the V$_\textrm{BD}$ values recorded at 20 $^{\circ}$C by the measurements shown in Figs.~\ref{fig:PadI_W3058} and~\ref{fig:PadI_W3081}. 
The total doping profile for the 20 $\mu\textrm{m}$-thick AC-LGAD in Fig.~\ref{LGADdoping} was reached by fine-tuning the depth of a typical $p^+$-layer doping profile to reproduce the measured value of V$_\textrm{BD}$ at 20 $^{\circ}$C in Fig.~\ref{fig:PadI_W3081}.
For the 50 $\mu\textrm{m}$-thick LGAD of Figure~\ref{LGADdoping50}, this required the introduction of an analytical $p^+$ layer that retained the key parameters (depth, $N_\textrm{p}$, and its location) of a SIMS-measured $p^+$-layer doping profile, since it was observed that the $p^+$ layer in the SIMS-profiles was not reproducing the measured V$_\textrm{BD}$ at 20 $^{\circ}$C. %by several hundreds of Volts. 
 Since it was not possible to adjust such a difference in V$_\textrm{BD}$ by fine-tuning the depth of the $p^+$ layer, as it was done with the 20 $\mu\textrm{m}$-thick AC-LGAD, the analytical $p^+$ layer was introduced with a Gaussian profile to reach agreement with the measurement.
 % The tuned % $p$-well peak dopings 
% values of $N_\textrm{p}$ are $6.65\times10^{16}$, $7.11\times10^{16}$ and $6.63\times10^{16}~\textrm{cm}^{-3}$ (for the $n^+$ layer option) for the 20- and the two 50-$\mu\textrm{m}$ thicknesses, respectively. 
%
%%%%%%%%%%%%%%%%%%%%%%%%%%%%%%%%%%%%%%%%%%%%%%%
%
\subsection{Simulated temperature evolution of breakdown voltages}
%%%%
\begin{figure}[h!]
    \centering
    % \subfloat[]{\includegraphics[width=.6\linewidth]{figures/TCAD_LGAD_50um_rampup_pad_all.png}}\\
    % \subfloat[]{\includegraphics[width=.6\linewidth]{figures/analysis/TCAD_IV_curves_Combined_LGAD_50um.pdf}}\\
    \subfloat[]{\includegraphics[width=.48\linewidth]{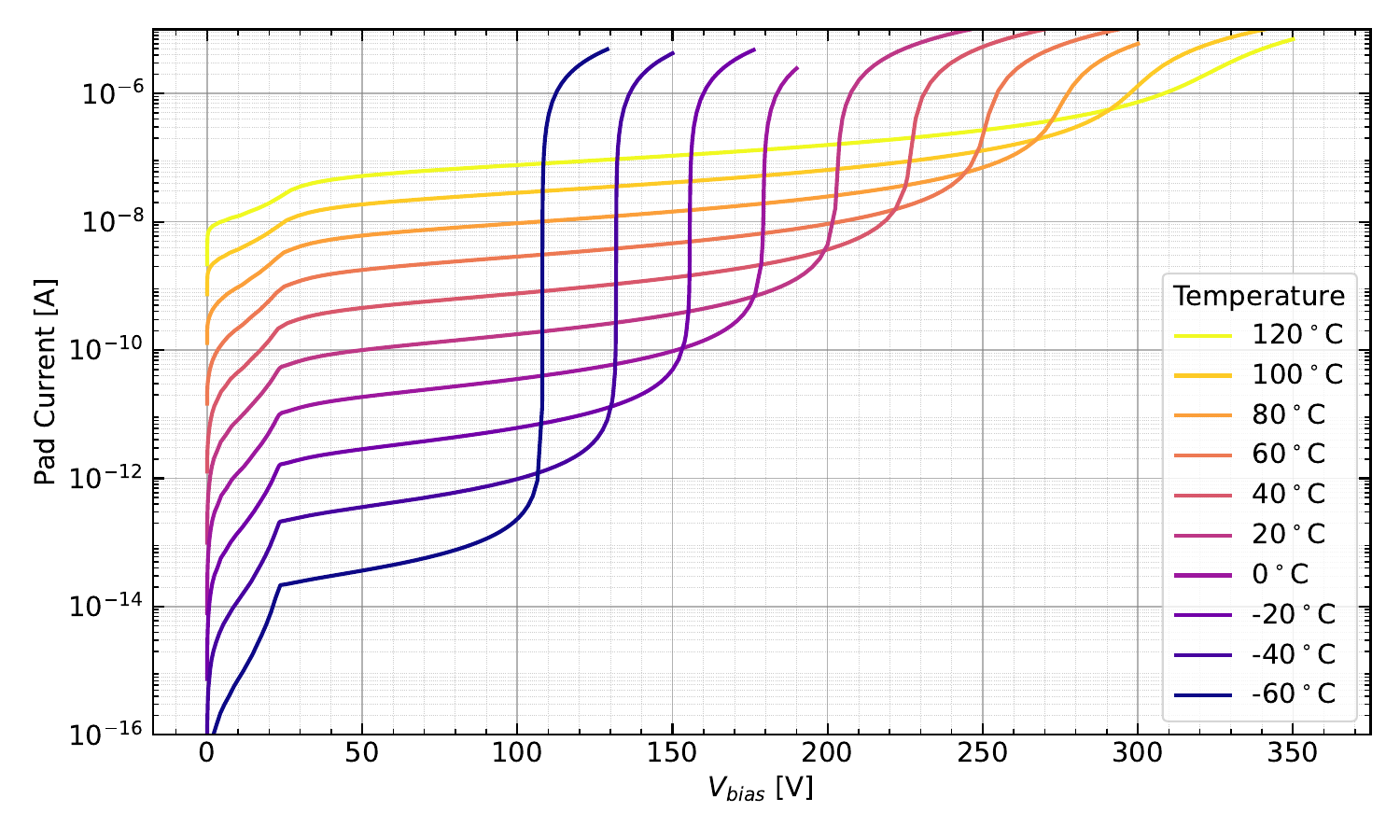}\label{TCAD_IV_50um}}
    \subfloat[]{\includegraphics[width=.48\linewidth]{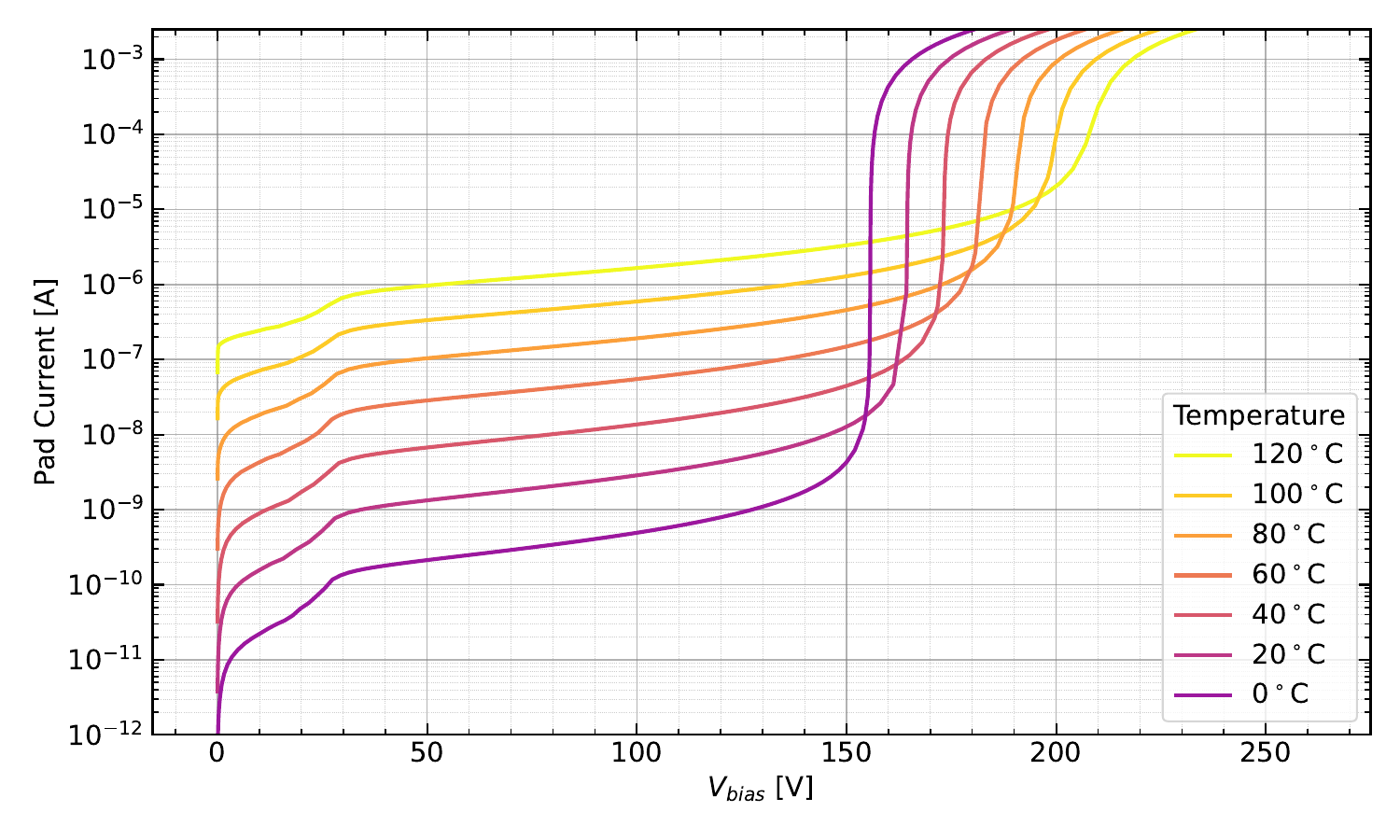}\label{TCAD_IV_20um}}
    \caption{\small % Simulated pad current as a function of bias voltage for different temperatures, (a) for $50~\mu\textrm{m}$ epi-layer thickness at temperatures from -60 to 120 $^{\circ}$C, corresponding to the LGAD W3058 in Figure~\ref{fig:PadI_W3045} (top), with dataset `$n^+$' % corresponds to corresponding to the high % peak doping at the $n^+$-layer doping in Figure~\ref{LGADdoping50} (middle), and for $20~\mu\textrm{m}$ epi-layer thickness at temperatures from 0 to 120 $^{\circ}$C, corresponding to the AC-LGAD W3081 in Figure~\ref{fig:PadI_W3081} (bottom).
    Simulated pad current as a function of bias voltage at different temperatures, for (a) $50~\mu\textrm{m}$ epi-layer thickness at temperatures from -60 to 120 $^{\circ}$C, corresponding to the LGAD W3058 in Figure~\ref{fig:PadI_W3058}, % with LGAD (solid lines) and AC-LGAD-like doping (dashed lines) at the % dataset `$n^+$' corresponding to the high $n^{+}$-layer % doping in Figure~\ref{LGADdoping50}, 
    and for (b) $20~\mu\textrm{m}$ epi-layer thickness at temperatures from 0 to 120 $^{\circ}$C, corresponding to the AC-LGAD W3081 in Figure~\ref{fig:PadI_W3081}.}
\label{fig:LGAD_TCAD_IV}
\end{figure}

%\begin{figure}[h!]
%\centering
% \subfloat[]{\includegraphics[width=3.0in]{pSample.png}\label{pSample}}\\%
% \hfil
% \subfloat[]{\includegraphics[width=.54\textwidth]{Rint_device2.png}\label{interPad}}\hspace{1mm}%
% \subfloat[]{\includegraphics[width=.45\textwidth]{HGCALdoping3a.png}\label{HGCALdoping}}%
%\subfloat[]{\includegraphics[width=.49\textwidth]{figures/LGAD_BNL_IV_50um2C.png}\label{IV_50um}}\hspace{1mm}%
%\subfloat[]{\includegraphics[width=.49\textwidth]{figures/LGAD_BNL_IV_20umC.png}\label{IV_20um}}%
%\caption{\small Simulated pad current as a function of bias voltage for different temperatures, (a) for $50~\mu\textrm{m}$ epi-layer thickness at temperatures from -60 to 120 $^{\circ}$C, corresponding to the LGAD W3058 in Figure~\ref{fig:PadI_W3045}, with dataset `$n^+$' % corresponds to 
%corresponding to the high % peak doping at the 
%$n^+$-layer doping in Figure~\ref{LGADdoping}; and (b) for $20~\mu\textrm{m}$ epi-layer thickness at temperatures from 0 to 120 $^{\circ}$C, corresponding to the AC-LGAD W3081 in Figure~\ref{fig:PadI_W3081}.
% \textcolor{red}{AT: convert temp from K to C in plots}
%}
%\label{LGAD_TCAD_IV}
%\end{figure}
%
% ...shown in 
Figures~\ref{TCAD_IV_50um} and~\ref{TCAD_IV_20um} present the IV-simulation results for the temperature ranges very similar to those in Figures~\ref{fig:PadI_W3058} and~\ref{fig:PadI_W3081}, respectively. The experimentally observed increases of leakage current and V$_\textrm{BD}$ with % $T$-independent $V_\textrm{fd}$, 
temperature were reproduced for both epi-thicknesses, i.e. 20 and 50 $ \mu\textrm{m}$. % The voltage where the active thickness of the device has become fully depleted $V_\textrm{fd}$, is reflected in the figures by the % $IV$-curves as a 
The distinct slope change in the IV-curves at the low-voltage region reflects the full depletion voltage, V$_\textrm{FD}$, where the active thickness of the device has become fully depleted of free charge carriers.
% The voltage where the active thickness of the device has become fully depleted $V_\textrm{fd}$, is reflected in the figures by the % $IV$-curves as a 
% distinct slope change in the $IV$-curves at the low-voltage region. 
The values of V$_\textrm{FD}$ are about 28 and 25 V for the 20 and 50 $\mu\textrm{m}$ thicknesses, respectively, % with $n$-resistive layer (AC-LGAD), respectively, and about 30 V for 50-$\mu\textrm{m}$ thickness with $n^+$ layer (LGAD) 
and remain constant % throughout the $T$-ranges 
for all temperatures as in the experimental results. The relative increase of currents as a function of temperature at V$_\textrm{FD}<$V$_{\mathrm{bias}}$<V$_\textrm{BD}$ % in Figs.~\ref{IV_50um} and~\ref{IV_20um} 
is in % close 
agreement with the measured increase for temperatures equal to or greater than 0 $^{\circ}$C and % essentially 
for the full temperature range, % \footnote{The unexpectedly high measured current at $+60~^{\circ}\textrm{C}$ in Fig.~\ref{fig:PadI_W3081} is not supported by the simulation.}, % \textcolor{red}{(AT: is this exception worth mentioning? or can you quantify the discrepancy? is there an explanation?)} ), 
respectively for Figs.~\ref{TCAD_IV_50um} and~\ref{TCAD_IV_20um}, following a power-law dependence. The disagreement between simulation and data at temperatures below 0 $^{\circ}$C for the 50~$\mu\textrm{m}$ thick LGAD can be explained by the limitation of the HP 4145B readout unit's sensitivity at currents less than 1 pA. Thus, the simulation provides a correction prediction for the measured leakage current in this temperature range. % in Figure~\ref{fig:IV_temp} (right)
Figure~\ref{fig:LGAD_TCAD_IV} also shows a clear dependence of the breakdown voltage on the epi-thickness, with the 20~$\mu\textrm{m}$ thickness having a lower breakdown voltage at the same temperature as the 50~$\mu\textrm{m}$ thickness, as shown in the data in Figs.~\ref{fig:PadI_W3058} and~\ref{fig:PadI_W3081}.
%%%%%%%%%%%%%%%%%%%%%%%%%%%%%%%%%%%%%%%%%%%%%%%

\begin{figure}[t!]
\centering
% \subfloat[]{\includegraphics[width=.495\textwidth]{LD_full.png}\label{LDfull}}\hspace{1mm}%
% \subfloat[]{
\includegraphics[width=.80\textwidth]{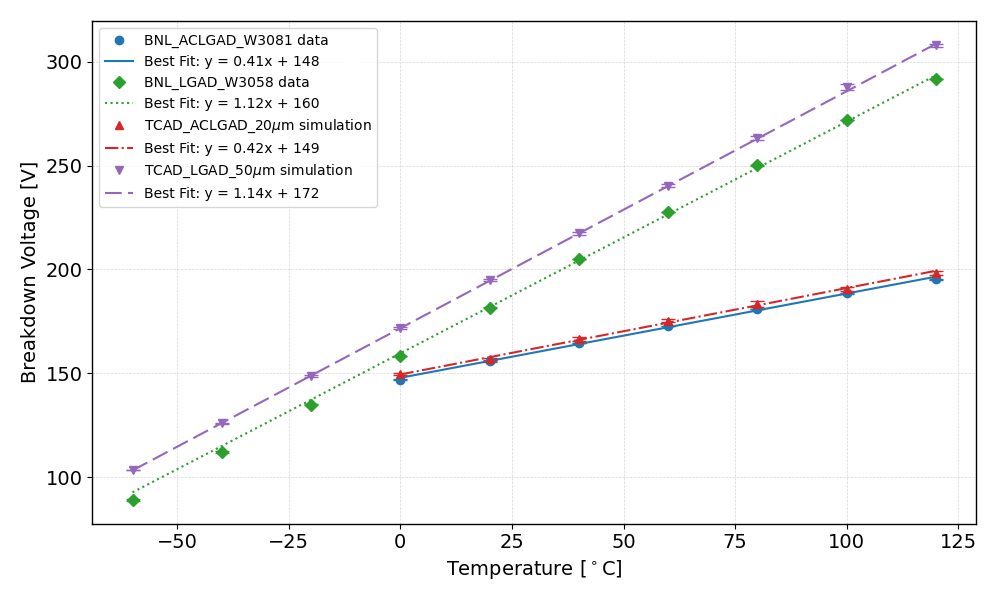}% \label{wafer}}\hspace{1mm}% \label{gds}%
% \subfloat[]{\includegraphics[width=.495\textwidth]{LD_full.png}\label{LDfull}}%
% \subfloat[]{\includegraphics[width=.4\textwidth]{HD_full.png}\label{HDfull}}\\% 
% \subfloat[]{\includegraphics[width=.4\textwidth]{HD_MGW.png}\label{HDmgw}}%
\caption{\small Measured and TCAD-simulated breakdown voltage as a function of temperature for the 20 and 50 $\mu\textrm{m}$-thick AC-LGAD W3081 and LGAD W3058, respectively. Simulated values of the breakdown voltage were extracted % identically to 
with the method introduced in Appendix~\ref{app:vBD-method}. The parameters of the linear fits for each dataset are displayed in the plot.}
% SQC $IV$-characterization grading of 300-$\upmu\textrm{m}$-thick production sensor `115732'. The analysis code grades the four requirements under `Global characteristics' Passed/Failed. The acceptance criteria for 200-$\upmu\textrm{m}$-thick LD-sensors are identical, while the expected depletion voltage in item two of Info-section is 130 V ($130\times1.5=195~\textrm{V}$).}
\label{VbdT_simMeas}
\end{figure}

The measured and simulated evolution of V$_\textrm{BD}$ with % as a function of 
temperature is presented in Fig.~\ref{VbdT_simMeas}. The simulated % The 
slope of V$_\textrm{BD}$ as a function of temperature % extracted from the simulations 
for the 20~$\mu\textrm{m}$-thick AC-LGAD displays considerable agreement with the experimentally extracted slope from Fig.~\ref{fig:temp_vs_breakdown}, i.e. $0.41\pm0.01$ and $0.42\pm0.01$ V/$^{\circ}$C in data and simulation, respectively. 
%%%%%%%%%%%%%%%%%%%%%%%%%%%%
% The two approaches for the simulation of the 50-$\mu\textrm{m}$-thick sensor with AC-LGAD and LGAD-like doping at the gain layer, show % slight but noticeable...
% slight difference in the 
% nonidentical slopes, i.e. $1.11\pm 0.01$ and $1.06\pm 0.02$ V/$^{\circ}$C for LGAD- and AC-LGAD-like doping respectively. This % can also be observed 
% is also evident in Figure~\ref{IV_50um}, where V$_\textrm{BD}$ values of the sensor with LGAD-like doping % has slightly lower currents and higher $V_\textrm{bd}$ values than the AC-LGAD at the same temperatures.
% display stronger temperature sensitivity than the AC-LGAD.
%%%%%%%%%%%%%%%%%%%%%%%%%%%
% Comparison with the % experimentally extracted slope...
% measured LGAD's slope of $1.12\pm 0.01$ V/$^{\circ}$C shows that the slopes in simulation are compatible to those in data. Thus, the substantially close agreement between the measured LGAD and the simulation with LGAD-like doping suggests that an accurate simulation of the dependence of the breakdown voltage as a function of temperature for LGADs and AC-LGADs require dedicated % device structures 
% gain layer tuning for each sensor type.
%%%%%%%%%%%%%%%%%%%%%%%%%%%%%%
The corresponding comparisons for the 50~$\mu\textrm{m}$-thick LGAD are $1.12\pm0.01$ and $1.14\pm0.01$ V/$^{\circ}$C for the measured and simulated cases, respectively; this shows that the slope in the simulation is compatible with the data. Thus, the substantially close agreement between the measured 50~$\mu\textrm{m}$-thick LGAD and the simulation suggests that the analytical doping profile with Gaussian decay in Figure~\ref{LGADdoping50} is a fair description of the real $p^+$ layer in the device.

% The results in Fig.~\ref{VbdT_simMeas} also confirm the experimental results shown in Fig. XX of an epi-layer thickness dependence of the rate of change (slope) of the breakdown voltage as a function of temperature, with a thicker epi-layer having a higher slope. This effect can be explained by the higher electric field at the breakdown voltage in a sensor with a thick epi-layer, as visible in Fig.~\ref{LGAD_TCAD_IV}.
The results in Fig.~\ref{VbdT_simMeas} also confirm the experimental results shown in Fig.~\ref{fig:temp_vs_breakdown} of an epi-layer thickness-dependence of the rate of change (slope) of the breakdown voltage as a function of temperature, with a thicker epi-layer having a higher slope. 
%\textcolor{red}{As a cross-check, the simulation was also performed for an LGAD and an AC-LGAD of the same thickness, i.e. 50~$\mu\textrm{m}$, and the slope change is small, i.e. $1.13\pm0.01$ and $1.3\pm0.01$ V/$^{\circ}$C, respectively, indicating that the slope is driven by the epi-layer thickness rather than the device processing.}
This is expected since a higher epi-layer thickness will require a higher V$_\textrm{bias}$ to reach the critical electric field for the avalanche onset. Thus, with these observations, the steepest slopes of the V$_\textrm{BD}$ dependence on temperature can be expected for LGADs with the thickest epi-layers.
%\textcolor{red}{The ratio of breakdown voltages as a function of temperature, shown in Fig. XX for data and simulation where the breakdown voltages are normalized to the value found at 20 $^{\circ}$C, is independent on the epi-layer thickness since the epi-thickness dependence cancel out on the ratio.}
%\textcolor{red}{AT: Add the plot of ratios of Vbd. }

% Accurate gain layer doping has a non-negligble role...
% After electrical SQC characterizations the acquired data was analyzed by {\it HGCAL Analysis Workflow}\footnote{https://gitlab.cern.ch/CLICdp/HGCAL/lcd\_hgcal\_analysisworkflows/} software and shared with HGCAL community in CERNBox cloud platform. Examples of the grading steps of a sensor's electrical performance are given in Figures~\ref{IV_acceptance} and~\ref{CV_acceptance}. None of the tested sensors failed the requirements for $CV$-performance.
%
%%%%%%%%%%%%%%%%%%%%%%%%%%%%%%%%%%%%%%%%%%%%%%%%%FIG. 3!!!
%% \begin{figure*}% [htb]
% \begin{comment}
%
%%%%%%%%%%%%%%%%%%%%%%%%%%%%%%%%%%%%%
\section{Conclusions}\label{sec:conclusions}

Two LGAD sensors and an AC-LGAD sensor fabricated by BNL were tested for weeks at rapidly changing temperature and humidity conditions, and the experimental results were compared to dedicated TCAD simulations. The results show a consistent response of leakage current as well as depletion, breakdown, and operating voltages as functions of temperature and humidity for all sensors. No deterioration of performance was observed in the sensors after weeks of thermal cycling. The new LGAD with an improved termination design and passivation shows lower leakage current than the old sensor design. For the three sensors, it is observed that the depletion voltage is independent of temperature and humidity, whereas the leakage current depends strongly on temperature and increases considerably above a relative humidity of about 50\%, equivalent to a sensor temperature within about 10 $^{\circ}$C of the dew point.
The breakdown voltage is independent of the humidity but depends linearly on temperature.
The rate of change of the breakdown voltage as a function of temperature depends on the active thickness of the epi-layer, with thicker epi-layers showing larger slopes.
The simulations with parameters tuned to reproduce the sensors' geometries, processing, and doping profiles as well as electrical properties are in agreement with experimental results. 

\appendix
\section{Method for the estimation of the breakdown voltage}
\label{app:vBD-method}% Preamble

\begin{figure}[htb]
   \centering
  \includegraphics[width=.8\linewidth]{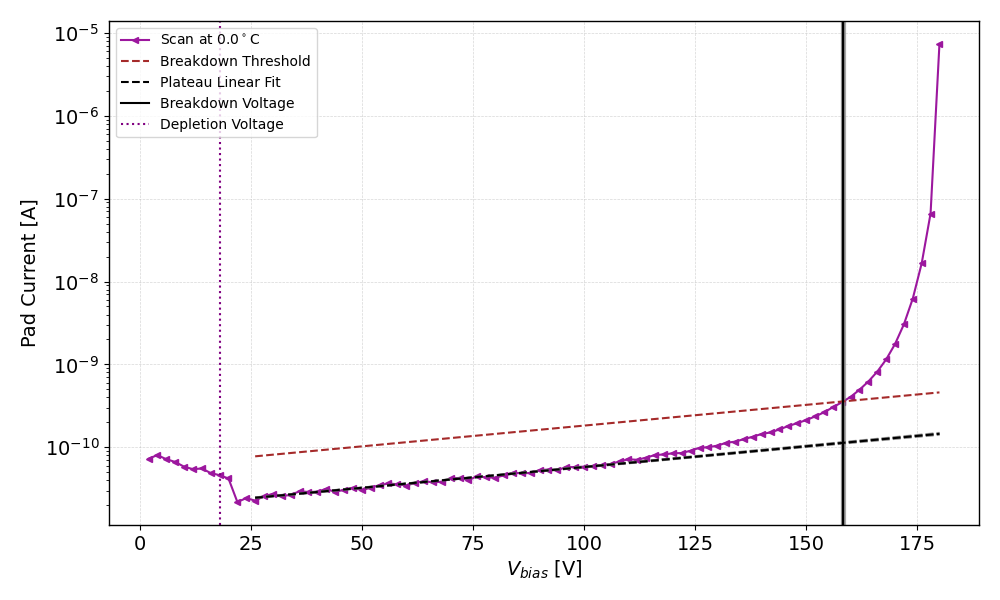}
   \caption{The breakdown voltage calculation method. The pad leakage current of LGAD W3058 is shown as a function of bias voltage for a single measurement at 0 $^{\circ}$C. The estimated depletion voltage is marked by the purple dashed vertical line. The fit obtained using RANSAC is shown by the dashed black line, and a number of other fits from the distribution are shown by the dashed gray lines (only partly visible because they overlap with the dashed gray line). The brown dashed line is I$^{\textrm{Pad}}_{\textrm{Th}}$ above the best fit. The vertical black line is at the selected V$_{\mathrm{BD}}$ and the vertical gray lines represent the breakdown voltage calculated from other random fits (the vertical gray lines are only partly visible because they overlap with the vertical black line).}
    \label{fig:breakdown_example}
\end{figure}

A persistent threshold method based on the RANdom SAmple Consensus (RANSAC) algorithm~\cite{RANSAC} is used to determine the V$_\textrm{BD}$ from IV measurements with the presence of noise and limited post-breakdown data points. This is graphically described in Figure \ref{fig:breakdown_example}.

The conventional approach~\cite{Vakili_2023} defines V$_\textrm{BD}$ as the intersection of linear fits to the plateau and post-breakdown regions, but it becomes unreliable when the latter contains too few data points. The method developed in this study instead identifies a sustained departure from the extrapolated plateau trend of the IV curve.  

The plateau region is defined as the first half of the data points above the estimated full depletion voltage. In each of $N$ trials, one-third of the plateau points of the IV scan, graphed with the current in logarithmic scale, is randomly selected and fit with a straight line. Each fit is evaluated over the full plateau region, and data points whose vertical distances from the fit line are smaller than a tolerance value $t$ in log-current scale is considered to be inliers. The fits with most inliers are selected, with ties resolved by the lowest mean squared error. Higher $N$ increases the probability of finding an optimal fit at the cost of compute time, and lower $t$ reduces the allowed deviation from fits and thus enhances accuracy at the cost of increased noise sensitivity. In this implementation, $N$ is set to 200 to achieve good fits with efficiency, while $t$ is set to 0.15 (in log-current scale) to balance fit accuracy and noise robustness. This random sampling and inlier voting process is known as RANSAC, a noise-robust regression method. The selected fit determines V$_\textrm{BD}$ as the lowest voltage at which the curve exceeds the fit line by a threshold of 0.5 (in log-current scale), defined as \textit{current threshold (I$^{\textrm{Pad}}_{\textrm{Th}}$)}, and remains above it at all higher measured voltages. This I$^{\textrm{Pad}}_{\textrm{Th}}$ threshold is chosen to minimize the uncertainty of V$_\textrm{BD}$ for all sensors under study. The spread of the V$_\textrm{BD}$ estimated from all 200 trials after removing extreme values is taken to compute the uncertainty.
\clearpage 
%%%
\section*{Acknowledgements}
The authors wish to thank their colleagues at Brookhaven National Laboratory:  Don Pinelli, Antonio Verderosa, Joe Pinz and Tim Kersten for sensor mounting as well as Wei Chen and Sean Robinson for sensor fabrication. This material is based upon work supported by the U.S. Department of Energy under grants DE-447 SC0012704, DE-SC443363, DE-SC426496 and DE-448 SC0020255 and the Brown University Undergraduate Teaching and Research Awards and the Brown Research Seed Award No.GR300352. This research used resources of the Center for Functional Nanomaterials, which is a U.S. DOE Office of Science Facility, at Brookhaven National Laboratory under Contract No. DE-SC0012704.
%%%%
%\newpage

% To print the credit authorship contribution details
\printcredits

%% Loading bibliography style file
% \bibliographystyle{model1-num-names}
% \bibliographystyle{cas-model2-names}

% Loading bibliography database
% \bibliography{cas-refs}

\bibliographystyle{elsarticle-num}
\bibliography{bib}

\begin{thebibliography}{10}
\expandafter\ifx\csname url\endcsname\relax
  \def\url#1{\texttt{#1}}\fi
\expandafter\ifx\csname urlprefix\endcsname\relax\def\urlprefix{URL }\fi
\expandafter\ifx\csname href\endcsname\relax
  \def\href#1#2{#2} \def\path#1{#1}\fi

\bibitem{hartmut}
H.~F.-W. Sadrozinski, et~al., Ultra-fast silicon detectors, Nucl. Inst. Meth. A 730 (2013) 226--231.
\newblock \href {https://doi.org/10.1016/j.nima.2013.06.033} {\path{doi:10.1016/j.nima.2013.06.033}}.

\bibitem{Giacomini_2019_LGAD}
G.~Giacomini, et~al., \href{http://dx.doi.org/10.1016/j.nima.2019.04.073}{Development of a technology for the fabrication of low-gain avalanche diodes at {BNL}}, Nucl. Inst. Meth. A 934 (2019) 52–57.
\newblock \href {https://doi.org/10.1016/j.nima.2019.04.073} {\path{doi:10.1016/j.nima.2019.04.073}}.
\newline\urlprefix\url{http://dx.doi.org/10.1016/j.nima.2019.04.073}

\bibitem{CMS:2667167}
{CMS Collaboration}, \href{https://cds.cern.ch/record/2667167}{{A MIP Timing Detector for the CMS Phase-2 Upgrade}}, Tech. Rep. CERN-LHCC-2019-003. CMS-TDR-020, CERN, Geneva (Mar 2019).
\newline\urlprefix\url{https://cds.cern.ch/record/2667167}

\bibitem{Collaboration:2623663}
{ATLAS Collaboration}, \href{http://cds.cern.ch/record/2623663}{{Technical Proposal: A High-Granularity Timing Detector for the ATLAS Phase-II Upgrade}}, Tech. Rep. CERN-LHCC-2018-023. LHCC-P-012, CERN, Geneva (Jun 2018).
\newline\urlprefix\url{http://cds.cern.ch/record/2623663}

\bibitem{AbdulKhalek:2021gbh}
R.~Abdul~Khalek, et~al., {Science Requirements and Detector Concepts for the Electron-Ion Collider}: {EIC Yellow Report}, Nucl. Phys. A 1026 (2022) 122447.
\newblock \href {http://arxiv.org/abs/2103.05419} {\path{arXiv:2103.05419}}, \href {https://doi.org/10.1016/j.nuclphysa.2022.122447} {\path{doi:10.1016/j.nuclphysa.2022.122447}}.

\bibitem{micronlgad}
N.~Moffat, et~al., Low gain avalanche detectors ({LGAD}) for particle physics and synchrotron applications, JINST 13 (2018) C~03014.
\newblock \href {https://doi.org/10.1088/1748-0221/13/03/C03014} {\path{doi:10.1088/1748-0221/13/03/C03014}}.

\bibitem{pioneercollaboration2022testingleptonflavoruniversality}
{PIONEER Collaboration}, \href{https://arxiv.org/abs/2203.05505}{Testing lepton flavor universality and ckm unitarity with rare pion decays in the pioneer experiment} (2022).
\newblock \href {http://arxiv.org/abs/2203.05505} {\path{arXiv:2203.05505}}.
\newline\urlprefix\url{https://arxiv.org/abs/2203.05505}

\bibitem{instruments5040040}
S.~M. Mazza, \href{https://www.mdpi.com/2410-390X/5/4/40}{{An LGAD-Based Full Active Target for the PIONEER Experiment}}, Instruments 5~(4) (2021).
\newblock \href {https://doi.org/10.3390/instruments5040040} {\path{doi:10.3390/instruments5040040}}.
\newline\urlprefix\url{https://www.mdpi.com/2410-390X/5/4/40}

\bibitem{Heller_2022}
R.~Heller, et~al., \href{http://dx.doi.org/10.1088/1748-0221/17/05/P05001}{Characterization of {BNL} and {HPK} {AC-LGAD} sensors with a 120 {GeV} proton beam}, JINST 17~(05) (2022) P05001.
\newblock \href {https://doi.org/10.1088/1748-0221/17/05/p05001} {\path{doi:10.1088/1748-0221/17/05/p05001}}.
\newline\urlprefix\url{http://dx.doi.org/10.1088/1748-0221/17/05/P05001}

\bibitem{ACLGADprocess}
G.~Giacomini, et~al., {Fabrication and Performance of AC-coupled LGADs}, JINST 14 (2019) P09004.
\newblock \href {https://doi.org/10.1088/1748-0221/14/09/P09004} {\path{doi:10.1088/1748-0221/14/09/P09004}}.

\bibitem{DUTTA2025170224}
I.~Dutta, et~al., \href{https://www.sciencedirect.com/science/article/pii/S0168900225000257}{{Results for pixel and strip centimeter-scale AC-LGAD sensors with a 120 GeV proton beam}}, Nucl. Inst. Meth. A 1072 (2025) 170224.
\newblock \href {https://doi.org/https://doi.org/10.1016/j.nima.2025.170224} {\path{doi:https://doi.org/10.1016/j.nima.2025.170224}}.
\newline\urlprefix\url{https://www.sciencedirect.com/science/article/pii/S0168900225000257}

\bibitem{Sze2007}
S.~M. Sze, et~al., Physics of Semiconductor Devices, 3rd Edition, Wiley-Interscience, Hoboken, NJ, 2007.

\bibitem{Chilingarov_2013}
A.~Chilingarov, \href{https://doi.org/10.1088/1748-0221/8/10/P10003}{Temperature dependence of the current generated in si bulk}, JINST 8~(10) (2013) P10003.
\newblock \href {https://doi.org/10.1088/1748-0221/8/10/P10003} {\path{doi:10.1088/1748-0221/8/10/P10003}}.
\newline\urlprefix\url{https://doi.org/10.1088/1748-0221/8/10/P10003}

\bibitem{ActivationEnergy}
V.~V.~N. Obreja, et~al., \href{https://onlinelibrary.wiley.com/doi/abs/10.1002/pssa.200925387}{Activation energy values from the temperature dependence of silicon pn junction reverse current and its origin}, Physica Status Solidi (a) 207~(5) (2010) 1252--1256.
\newblock \href {http://arxiv.org/abs/https://onlinelibrary.wiley.com/doi/pdf/10.1002/pssa.200925387} {\path{arXiv:https://onlinelibrary.wiley.com/doi/pdf/10.1002/pssa.200925387}}, \href {https://doi.org/https://doi.org/10.1002/pssa.200925387} {\path{doi:https://doi.org/10.1002/pssa.200925387}}.
\newline\urlprefix\url{https://onlinelibrary.wiley.com/doi/abs/10.1002/pssa.200925387}

\bibitem{UniBo2005}
S.~Reggiani, et~al., {Measurement and modeling of the electron impact-ionization coefficient in silicon up to very high temperatures}, IEEE Trans. Electron Dev. 52 (2005) 2290--2299.
\newblock \href {https://doi.org/10.1109/TED.2005.856807} {\path{doi:10.1109/TED.2005.856807}}.

\bibitem{Massey2006}
D.~Massey, et~al., {Temperature Dependence of Impact Ionization in Submicrometer Silicon Devices}, IEEE Trans. Electron Dev. 53 (2006) 2328--2334.
\newblock \href {https://doi.org/10.1109/TED.2006.881010} {\path{doi:10.1109/TED.2006.881010}}.

\bibitem{Poehlsen2013time}
T.~Poehlsen, et~al., Time dependence of charge losses at the {Si}--{SiO}$_2$ interface in {p$^+$n}-silicon strip sensors, Nucl. Inst. Meth. A 731 (2013) 172--179, arXiv:1305.0398.
\newblock \href {https://doi.org/10.1016/j.nima.2013.03.023} {\path{doi:10.1016/j.nima.2013.03.023}}.

\bibitem{Poehlsen2013charges}
T.~Poehlsen, et~al., Charge losses in segmented silicon sensors at the {Si}--{SiO}$_2$ interface, Nucl. Inst. Meth. A 700 (2013) 22--39.
\newblock \href {https://doi.org/10.1016/j.nima.2012.10.063} {\path{doi:10.1016/j.nima.2012.10.063}}.

\bibitem{Ranjan2002}
K.~Ranjan, et~al., Analysis and comparison of the breakdown performance of semi-insulator and dielectric passivated {Si} strip detectors, Nucl. Inst. Meth. A 492~(3) (2002) 536--545.
\newblock \href {https://doi.org/10.1016/S0168-9002(02)01609-1} {\path{doi:10.1016/S0168-9002(02)01609-1}}.

\bibitem{Karrevula2023}
V.~R. Karrevula, et~al., Effect of pre-adsorbed moisture and humidity on {I--V} characteristics of {Si PIN} diode, Nucl. Inst. Meth. A 1047 (2023) 167832.
\newblock \href {https://doi.org/10.1016/j.nima.2022.167832} {\path{doi:10.1016/j.nima.2022.167832}}.

\bibitem{instr:keithley2410}
Keithley Instruments, Inc., Series 2400 SourceMeter: User’s Manual (September 2011).

\bibitem{instr:HP4145b}
Hewlett-Packard Co., {Operational Manual HP 4145B Semiconductor Parameter Analyzer}, {HP 4145B} (April 1989).

\bibitem{Keithley6482}
Keithley, Tektronix, {Model 6482 Dual-Channel Picoammeter/Voltage Source}.

\bibitem{sensirion2011sht7x}
Sensirion AG, \href{www.sensirion.com}{Datasheet SHT7x (SHT71, SHT75) Humidity and Temperature Sensor IC}, version 5 (December 2011).
\newline\urlprefix\url{www.sensirion.com}

\bibitem{arduino_uno}
{Arduino}, {Arduino Uno Rev3}, \url{https://store.arduino.cc/products/arduino-uno-rev3}, accessed: 2026-06-25.

\bibitem{climateChamber:TJR}
Thermal Product Solutions, Operation and Maintenance Manual (January 2018).

\bibitem{vO1970}
R.~V. Overstraeten, et~al., {Measurement of the ionization rates in diffused silicon p-n junctions}, Solid-State Electronics 13 (1970) 583--608.
\newblock \href {https://doi.org/10.1016/0038-1101(70)90139-5} {\path{doi:10.1016/0038-1101(70)90139-5}}.

\bibitem{Okuto1975}
Y.~Okuto, et~al., {Threshold energy effect on avalanche breakdown voltage in semiconductor junctions}, Solid-State Electronics 18 (1975) 161--168.
\newblock \href {https://doi.org/10.1016/0038-1101(75)90099-4} {\path{doi:10.1016/0038-1101(75)90099-4}}.

\bibitem{Lackner1991}
T.~Lackner, {Avalanche multiplication in semiconductors: A modification of Chynoweth's law}, Solid-State Electronics 34 (1991) 33--42.
\newblock \href {https://doi.org/10.1016/0038-1101(91)90197-7} {\path{doi:10.1016/0038-1101(91)90197-7}}.

\bibitem{UniBo2_2004}
S.~Reggiani, et~al., {Experimental extraction of the electron impact-ionization coefficient at large operating temperatures}, {IEDM Technical Digest} (2004) 407--410\href {https://doi.org/10.1109/IEDM.2004.1419171} {\path{doi:10.1109/IEDM.2004.1419171}}.

\bibitem{Peltola2026_DRD3}
T.~Peltola, et~al., {TCAD simulations of AC-LGAD sensors: Impact ionization models \& influence of temperature}, \href{https://indico.cern.ch/event/1634258/contributions/7138033/}{5th DRD3 week on Solid State Detectors R\&D} (June 2026).

\bibitem{RANSAC}
R.~Bolles, et~al., \href{https://dl.acm.org/doi/10.1145/358669.358692}{Random sample consensus: a paradigm for model fitting with applications to image analysis and automated cartography}, Communications of the ACM 24 (1981) 381--395.
\newblock \href {https://doi.org/10.1145/358669.358692} {\path{doi:10.1145/358669.358692}}.
\newline\urlprefix\url{https://dl.acm.org/doi/10.1145/358669.358692}

\bibitem{Vakili_2023}
A.~Vakili, et~al., \href{https://doi.org/10.1088/1748-0221/18/07/P07052}{Analysis of the performance of low gain avalanche diodes for future particle detectors}, JINST 18~(07) (2023) P07052.
\newblock \href {https://doi.org/10.1088/1748-0221/18/07/P07052} {\path{doi:10.1088/1748-0221/18/07/P07052}}.
\newline\urlprefix\url{https://doi.org/10.1088/1748-0221/18/07/P07052}

\end{thebibliography}

% Biography
%\bio{}
% Here goes the biography details.
%\endbio

%\bio{pic1}
% Here goes the biography details.
%\endbio

\end{document}